\documentclass[preprint]{elsarticle}

\usepackage{newa4}      
\usepackage{latexsym}   
\usepackage{amsmath}    
\usepackage{amssymb}    
\usepackage{amsfonts}   
\usepackage{accents}
\usepackage{graphicx}
\usepackage{proof}

\newtheorem{thm}{Theorem}
\newtheorem{lem}[thm]{Lemma}
\newtheorem{cor}[thm]{Corollary}
\newtheorem{pn}[thm]{Proposition}
\newdefinition{dn}{Definition}
\newproof{pf}{Proof}

\newcommand{\apmap}[3]{\mbox{$#1 \colon #2 \Vdash #3$}}
\newcommand{\dda}{\mathord{\mbox{\makebox[0pt][l]{\raisebox{-.4ex}
                           {$\downarrow$}}$\downarrow$\,}}}
\newcommand{\fsubset}{\subseteq_\mathrm{fin}}
\newcommand{\fun}[3]{\mbox{$#1 \colon #2 \rightarrow #3$}}
\newcommand{\low}{\mathopen\downarrow\,}
\newcommand{\pair}[1]{\langle #1 \rangle}
\newcommand{\set}[2]{\mbox{$\{\,#1 \mid #2 \,\}$}}
\newcommand{\up}{\mathopen\uparrow}
\newcommand{\Dashv}{\mathrel{\reflectbox{$\vDash$}}}
\newcommand{\xsim}{\mathrel{\underset{\bigcup X_{i}}{\sim}}}
\newcommand{\hxsim}{\mathrel{\widehat{\underset{\bigcup X_{i}}{\sim}}}} 
\newcommand{\newand}{\mathop{\&}}

\def\con{\mathop{\mathstrut\rm Con}\nolimits}
\def\Con{\mathop{\mathstrut\rm CON}\nolimits}
\def\dx{\mathop{\mathstrut\rm dp}}
\def\id{\mathop{\mathstrut\rm id}\nolimits}
\def\Id{\mathop{\mathstrut\rm Id}}
\def\ip{\mathop{\mathstrut\rm IP}}
\def\pr{\mathop{\mathstrut\rm pr}\nolimits}
\def\sp{\mathop{\mathstrut\rm sp}}
\def\pd{\mathop{\mathstrut\rm pd}}
\def\st{\mathop{\mathstrut\rm st}}
\def\svee{\mathop{\mathstrut \dot{\vee}}}
\def\scup{\mathop{\mathstrut \dot{\sqcup}}}
\def\dvee{\mathop{\mathstrut \accentset{\bullet}\bigvee}}
\def\sqdvee{\mathop{\mathstrut \accentset{\bullet}\bigsqcup}}

\newcommand{\AAA}{\mathcal{A}}
\newcommand{\BBB}{\mathcal{B}}
\newcommand{\CCC}{\mathcal{C}}
\newcommand{\DDD}{\mathcal{D}}
\newcommand{\EEE}{\mathcal{E}}
\newcommand{\FFF}{\mathcal{F}}
\newcommand{\III}{\mathcal{I}}
\newcommand{\LLL}{\mathcal{L}}
\newcommand{\PPP}{\mathcal{P}}
\newcommand{\QQQ}{\mathcal{Q}}
\newcommand{\RRR}{\mathcal{R}}
\newcommand{\SSS}{\mathcal{S}}
\newcommand{\TTT}{\mathcal{T}}

\newcommand{\AAb}{\mathbb{A}}
\newcommand{\BB}{\mathbb{B}}
\newcommand{\CC}{\mathbb{C}}
\newcommand{\EE}{\mathbb{E}}
\newcommand{\FF}{\mathbb{F}}
\newcommand{\GG}{\mathbb{G}}
\newcommand{\LL}{\mathbb{L}}
\newcommand{\SSb}{\mathbb{S}}

\newcommand{\bA}{\mathbf{A}}
\newcommand{\bP}{\mathbf{P}}
\newcommand{\bQ}{\mathbf{Q}}
\newcommand{\bT}{\mathbf{T}}
\newcommand{\bx}{\mathbf{x}}
\newcommand{\bt}{\mathbf{t}}

\begin{document}

\title{Generalised Information Systems Capture L-Domains\tnoteref{t1}}
\tnotetext[t1]{The research leading to these results has received funding from the People Programme (Marie Curie Actions) of the European Union's Seventh Framework Programme FP7/2007-2013/ under REA grant agreement no. PIRSES-GA-2013-612638-CORCON.}
\author[1]{Dieter Spreen}
\ead{spreen@math.uni-siegen.de}
\address[1]{Department of Mathematics, University of Siegen, 57068 Siegen, Germany\\
and\\
Department of Decision Sciences, University of South Africa,
P.O. Box 392, 0003 Pretoria, South Africa}

\begin{abstract}
A generalisation of Scott's information systems~\cite{sco82} is presented that captures exactly all L-domains. The global consistency predicate in Scott's definition is relativised in such a way that there is a consistency predicate for each atomic proposition (token) saying which finite sets of such statements express information that is consistent with the given statement.

It is shown that the states of such generalised information systems form an L-domain, and that each L-domain can be generated in this way, up to isomorphism. Moreover, the equivalence of the category of generalised information systems with the category of L-domains is derived. 
In addition, it will be seen that from every generalised information system capturing an algebraic bounded-complete domain a corresponding Scott information system can be obtained in an easy and natural way, and vice versa; similarly for Hoofman's continuous information systems~\cite{ho93} and the continuous bounded-complete domains captured by them; for Chen and Jung's disjunctive propositional logic~\cite{cj06} and algebraic L-domains (as well as for Wang and Li's~\cite{wll20} finitary version and Lawson-compact algebraic L-domains); and for Wang and Li's conjunctive sequent calculi~\cite{wlbc20} and proper continuous bounded-complete domains. The proofs always contain syntactic translations between the logical calculi involved.
\end{abstract}


\date{}
\maketitle

 \tableofcontents 

\section{Introduction}\label{sec-intro}

In 1982, in his seminal paper \cite{sco82}, Dana Scott introduced information systems as a logic-based approach to domain theory. An information system consists of a set of tokens to be thought of as atomic statements about a computational process, a consistency predicate telling us which finite sets of such statements contain consistent information, and an entailment relation saying what atomic statements are entailed by which consistent sets of these. Theories  of such a logic, also called states, i.e.\ finitely consistent and entailment-closed sets of atomic statements, form an algebraic bounded-complete domain with respect to set inclusion, and, conversely, every such domain can be obtained in this way, up to isomorphism. This gives Scott's idea that domain elements represent information about states of a computation a precise mathematical meaning.

The role of bounded completeness becomes also clear in this context: States represent consistent information. So, any finite collection of substates must contain consistent information as well, and this fact is witnessed by any of its upper bounds.

Whereas in Scott's approach the consistency witnesses are hidden, in this paper we present an approach that makes them explicit. This allows to consider the more general situation in which there is no longer a uniform global consistency predicate. Instead there is a consistency predicate for each atomic statement telling us which finite sets of atomic statements express information that is consistent with the given statement. As it turns out the theories, or states, of such a more general information system form an L-domain, and, up to isomorphism, each L-domain can be obtained in this way. 

Since every token in the just delineated kind of information system has its own consistency predicate, we can also think of each such system as a family of logics, or a Kripke frame. 

L-domains were independently introduced by Coquand \cite{co89} and Jung \cite{ju89}. As was shown by Jung \cite{ju89,ju90}, they form one of the two maximal Cartesian closed full subcategories of the category of continuous domains with Scott continuous functions.

Here, we show that the category of generalized information systems and approximable mappings is equivalent to the category of L-domains. Similar results are derived for algebraic L-domains and bounded-complete domains. In both cases the corresponding generalised information systems satisfy just one additional condition.

Note that a logic-oriented approach to L-domains has also been presented by Zhang \cite{zh92a}. However, the representation considered in that paper is motivated by Gentzen-style proof systems and therefore differs from Scott's original approach. Moreover, only algebraic L-domains are captured and the function space construction is not considered. Chen and Jung \cite{cj06} developed a logic for describing algebraic L-domains following Abramsky's Domain Theory in Logical Form approach \cite{ab91}. The logic is an infinitary propositional logic allowing the formation of infinitely long disjunctions satisfying a disjointness property. Derivations are based on a Gentzen-style sequent calculus. A similar logic allowing only finite disjunctions is used by Wang and Li~\cite{wll20} for Lawson-compact algebraic L-domains. As shown by the same authors~\cite{wlbc20}, propositional logic with a constant for truth and conjunction as the only logical connective represents proper bounded-complete continuous domains.

A Scott-style logic-oriented approach capturing general L-domains was introduced by the present author in \cite{sp12}. However, as in the other approaches, consistency witnesses were hidden. Each consistent set was required to contain its witness, which led to the unsatisfying situation that subsets of consistent sets needed not be consistent again. The problem is eliminated in the present approach. All requirements now have a clear  logical meaning.

The approach in \cite{sp12} is based on a general type of information systems capturing all continuous domains~\cite{sxm08}. Wu et al.~\cite{wgl16} presented another extension of the conditions in \cite{sxm08}, as well leading to a representation of all L-domains.

As was shown in \cite{sxm08}, all important subclasses of these domains can be characterised by suitable additional axioms, only the case of L-domains was left open. So, in particular, bounded-complete continuous domains were captured. However, as was shown by Huang et al.~\cite{hzl15}, Hoofman's continuous information systems~\cite{ho93}, though doing the same job, could not be subsumed as special case.  The authors managed to simplify the axiom system in \cite{sxm08} so that now all types of information studied in  the literature before turn out to be special cases of the simplified kind of information systems. 

Scott's original motivation for the introduction of information systems was to provide a more concrete approach to (abstract) domain theory. Therefore, he presented information system analogues of the domain constructions usually needed in giving a denotational programming language semantics. Especially, the construction of exponents requires special attention in our case. It is the topic of a separate paper~\cite{sp?}.

The present paper is organized as follows: Section~\ref{sec-dom} contains basic definitions and results from domain theory. In Section~\ref{sec-infosys} generalised information systems are considered: the concepts of information frames and information systems with witnesses are introduced and their equivalence is derived. Moreover, it is shown that the states of an information system with witnesses form an L-domain with respect to set inclusion, and that---up to isomorphism---every L-domain can be generated  that way. The special cases of algebraic L-domains and bounded-complete domains are considered as well. 
Approximable mappings between information systems with witnesses are defined in Section~\ref{sec-appmp} and the equivalence between the category of such information systems and mappings and the category of L-domains is shown. 

Scott's information systems are known to represent exactly the algebraic bounded-com\-plete domains. The notion has been generalised by Hoofmann~\cite{ho93} to capture all continuous bounded-complete domains. In Section~\ref{sec-rel} two classes of information systems with witnesses are considered representing algebraic and continuous bounded-complete domains, respectively, and it is shown that one can pass in an easy and natural way from information systems with witnesses of this kind to Scott and/or Hoofmann information systems in such a way that (up to isomorphism) the same domains are represented, and vice versa.

In the same way, in Section~\ref{sec-domlog}, the relationship of other domain logics capturing L-domains, or subclasses thereof, to information systems with witnesses is investigated. We show that disjunctive propositional logic studied by Chen and Jung~\cite{cj06} induces an information system with witnesses such that both generate isomorphic domains, and vice versa; similarly for the finitary version investigated by Wang and Li~\cite{wll20}. In the same way, conjunctive sequent calculi, as studied by Wang and Li in \cite{wlbc20}, will be seen to define information systems with witnesses, and vice versa, again with isomorphic domains. In both Sections, \ref{sec-rel} and \ref{sec-domlog}, the proofs contain purely syntactic translations between the logical calculi involved.

The paper finishes with a Conclusion.
First results of this research have been presented at the Workshop on Domains XI, Paris, 2014.

\section{Domains: basic definitions and results}\label{sec-dom} 

For any set $A$, we write $X \fsubset A$ to mean that $X$ is finite subset of $A$. The collection of all subsets of $A$ will be denoted by $\PPP(A)$ and that of all finite subsets by $\PPP_f(A)$.

Let $(D, \sqsubseteq)$ be a poset. $D$ is \emph{pointed} if it contains a least element $\bot$. For an element $x \in D$, $\low x$ denotes the principal ideal generated by $x$, i.e., $\low x = \set{y \in D}{y \sqsubseteq x}$. A subset $S$ of $D$ is called \emph{consistent} if it has an upper bound. $S$ is \emph{directed}, if it is nonempty and every pair of elements in $S$ has an upper bound in $S$. $D$ is a \emph{directed-complete partial order} (\emph{dcpo}), if every directed subset $S$ of $D$ has a least upper bound $\bigsqcup S$ in $D$, and $D$ is \emph{bounded-complete} if every consistent subset of $D$ has a least upper bound in $D$.

Assume that $x, y$ are elements of a poset $D$. Then $x$ is said to \emph{approximate} $y$, written $x \ll y$, if for any directed subset $S$ of $D$ the least upper bound of which exists in $D$, the relation $y \sqsubseteq \bigsqcup S$ always implies the existence of some $u \in S$ with $x \sqsubseteq u$. Moreover, $x$ is \emph{compact} if $x \ll x$. A subset $B$ of $D$ is a \emph{basis} of $D$, if for each $x \in D$ the set $\dda\!_B x = \set{u \in B}{u \ll x}$ contains a directed subset with least upper bound $x$. Note that the set of all compact elements of $D$ is included in every basis of $D$.  A directed-complete partial order $D$ is said to be \emph{continuous} (or a \emph{domain}) if it has a basis and it is called \emph{algebraic} (or an \emph{algebraic domain}) if its compact elements form a basis. A pointed bounded-complete domain is called \emph{bc-domain}. Standard references for domain theory and its applications are \cite{gs, gu, aj, dom, ac, gie}.

\begin{lem}\label{lem-preordprop}
In a poset $D$ the following statements hold for all $x, y, z \in D$: \begin{enumerate}
\item\label{lem-preordprop-0} The approximation relation $\ll$ is transitive.
\item\label{lem-preordprop-1} $x \ll y \Longrightarrow x \sqsubseteq y$.
\item\label{lem-preordprop-2} $x \ll y \sqsubseteq z \Longrightarrow x \ll z$.
\item\label{lem-preordprop-4} If $D$ has a least element $\bot$, then $\bot \ll x$.
\item\label{lem-preordprop-3} If $F \subseteq \low x \cap \low y$ such that the least upper bounds $\bigsqcup^x F$ and $\bigsqcup^y F$, respectively, exist relative to $\low x$ and $\low y$, then 
\[
x, y \sqsubseteq z \Longrightarrow \bigsqcup\nolimits^x F = \bigsqcup\nolimits^y F.
\]
\item\label{lem_preordprop-5} If $D$ is a continuous domain with basis $B$, and $M \fsubset D$, then
\[
M \ll x \Longrightarrow (\exists v \in B) M \ll v \ll x,
\]
where $M \ll x$ means that $m \ll x$, for all $m \in M$.

\end{enumerate}
\end{lem}

Property~\ref{lem_preordprop-5} is known as the \emph{interpolation law}.

\begin{dn}
Let $D$ and $D'$ be posets. A function $\fun{f}{D}{D'}$ is \emph{Scott continuous} if it is monotone and for any directed subset $S$ of $D$ with existing least upper bound,
\[
\bigsqcup f(S) = f(\bigsqcup S).
\]
\end{dn}

With respect to the pointwise order the set $[D \to D']$ of all Scott continuous functions between two dcpo's $D$ and $D'$ is a dcpo again. Observe that it need not be continuous even if $D$ and $D'$ are. This is the case, however, if $D'$ is an L-domain \cite{aj}.

\begin{dn}
A pointed\footnote{Note that in \cite{gie} pointedness is not required.} domain $D$ is an \emph{L-domain}, if each pair $x, y \in D$ bounded above by $z \in D$ has a least upper bound $x \sqcup^z y$ in $\low z$.
\end{dn}

Obviously, every bc-domain is an L-domain.
As has been shown by Jung \cite{ju89,ju90}, the category $\mathbf{L}$ of L-domains is one of the two maximal Cartesian closed full subcategories of the category $\mathbf{CONT_\perp}$ of pointed domains and Scott continuous maps. The same holds for the category $\mathbf{aL}$ of algebraic L-domains with respect to the category $\mathbf{ALG_\perp}$ of pointed algebraic domains.  The one-point domain is the terminal object in these categories and the categorical product $D \times E$  of two domains $D$ and $E$ is the Cartesian product of the underlying sets  ordered coordinatewise.

\section{Generalised information systems}\label{sec-infosys}

In this section, the ideas outlined in the introduction are made precise: We introduce two---equivalent---generalisations of information systems and study their relationship with L-do\-mains. First,  information frames will be considered.

An information frame consists of a Kripke frame (A, R), the nodes of which are also called tokens. Associated with each node $i \in A$ is a consistency predicate $\con_i$ classifying the finite sets of tokens which are consistent with respect to node $i$, and an entailment relation $\vdash_i$ between $i$-consistent sets and tokens.

The conditions that have to be satisfied are grouped. There are requirements which consistency predicate and entailment relation of each single node have to meet, and which are well known from Scott's information systems. In addition, we find conditions that specify their interplay for nodes related to each other by the accessibility relation. 

\begin{dn}\label{dn-infofr}
Let $A$ be a set, $R$ be a binary relation on $A$, $\bt \in A$, $(\con_i)_{i \in A}$ be a family of subsets of $\PPP_f(A)$, and $(\vdash_i)_{i \in A}$ be a family of relations $\vdash_i  \subseteq \con_i \times A$. Then $\AAA = (A, R, (\con_i)_{i \in A}, (\vdash_i)_{i \in A}, \bt)$ is an \emph{information frame} if the following conditions hold, for all $i, j, a \in A$ and all finite subsets $X, Y$ of $A$:
\begin{enumerate}
\item\label{dn-infofr-1}
$\{i\} \in \con_i$

\item\label{dn-infofr-2}
$Y \subseteq X \wedge X \in \con_i \Rightarrow Y \in \con_i$

\item\label{dn-infofr-3}
$\emptyset \vdash_i \bt$

\hspace{-\leftmargin}
and, defining $X \vdash_i Y$ to mean that $X \vdash_i b$, for all $b \in Y$,

\item\label{dn-infofr-4}
$X \in \con_i\mbox{} \wedge X \vdash_i Y \Rightarrow Y \in \con_i$

\item\label{dn-infofr-5}
$X, Y \in \con_i\mbox{} \wedge Y \supseteq X \wedge X \vdash_i a \Rightarrow Y \vdash_i a$

\item\label{dn-infofr-6}
$X \in \con_i\mbox{} \wedge X \vdash_i Y \wedge Y \vdash_i a \Rightarrow X \vdash_i a$

\item\label{dn-infofr-6+}
$X \in \con_i\mbox{} \wedge X \vdash_i a \Rightarrow (\exists Z \in \con_i) X \vdash_i Z \wedge Z \vdash_i a$

\item\label{dn-infofr-7}
$i R j \Rightarrow \con_i \subseteq \con_j$

\item\label{dn-infofr-8}
$\{i\} \in \con_j \Rightarrow i R j$.

\item\label{dn-infofr-9}
$i R j \wedge X \in \con_i\mbox{} \wedge X \vdash_i a \Rightarrow X \vdash_j a$

\item\label{dn-infofr-10}
$i R j \wedge X \in \con_i\mbox{} \wedge X \vdash_j a \Rightarrow X \vdash_i a$

\item\label{dn-infofr-11}
$X \vdash_i Y \Rightarrow (\exists e \in A) X \vdash_i e \wedge Y \in \con_e$.

\end{enumerate}
\end{dn}

All requirements are very natural:  Each token witnesses its own consistency (\ref{dn-infofr-1}). If the consistency of some set is witnessed by $i$, the same holds for all of its subsets (\ref{dn-infofr-2}).  $\bt$ is entailed by any set of information and in every node, i.e., it represents global truth (\ref{dn-infofr-3}). By (\ref{dn-infofr-4}) each entailment relation preserves consistency. If a set $X$ entails $a$, so does any bigger set (\ref{dn-infofr-5}). Entailment is idempotent (\ref{dn-infofr-6},\ref{dn-infofr-6+}). In particular it is transitive.  Consistency and entailment are preserved when moving from a node $i$ to its accessible neighbour $j$ (\ref{dn-infofr-7},\ref{dn-infofr-9}). Moreover, entailment is conservative: what is $j$-entailed from an $i$-consistent set is already $i$-entailed (\ref{dn-infofr-10}). Condition~(\ref{dn-infofr-11}), finally, states an interpolation property strengthening (\ref{dn-infofr-4}).

\begin{lem}\label{lem-eq4}
Let $A$ be a set, $R$ be a binary relation on $A$, $\bt \in A$, $(\con_i)_{i \in A}$ be a family of subsets of $\PPP_f(A)$, and $(\vdash_i)_{i \in A}$ be a family of relations $\vdash_i  \subseteq \con_i \times A$ such that Axioms~\ref{dn-infofr}(\ref{dn-infofr-7},\ref{dn-infofr-8}) hold. Then the following two statements hold:
\begin{enumerate}
\item\label{lem-eq4-1} If Axiom~\ref{dn-infofr}(\ref{dn-infofr-4}) is satisfied, then for all $i, j \in A$ and all $X \in \con_i$,
\begin{equation}\label{eq-dashR}
 X \vdash_i j \Rightarrow j R i.
\end{equation}

\item\label{lem-eq4-2} If Condition~(\ref{eq-dashR}) and Axiom~\ref{dn-infofr}(\ref{dn-infofr-11}) are satisfied, so is Axiom~
\ref{dn-infofr}(\ref{dn-infofr-4}). 
\end{enumerate}
\end{lem}
\begin{pf}
(\ref{lem-eq4-1}) If $X \vdash_i j$, then $\{ j \} \in \con_i$, by Axiom~\ref{dn-infofr}(\ref{dn-infofr-4}), and thus $j R i$, because of Condition~\ref{dn-infofr}(\ref{dn-infofr-8}).

(\ref{lem-eq4-2}) Assume that $X \vdash_i Y$. Because of Axiom~\ref{dn-infofr}(\ref{dn-infofr-11}) there is some $j \in A$ such that $X \vdash_i j$ and $Y \in \con_j$. With (\ref{eq-dashR}) we obtain that $ j R i$ and hence with Condition~\ref{dn-infofr}(\ref{dn-infofr-7}) that $Y \in \con_i$.
\end{pf}

By Axiom~\ref{dn-infofr}(\ref{dn-infofr-6+}) the entailment relation of each node of an information frame satisfies an interpolation condition. As we will see now, the frame also has a global interpolation property.

\begin{lem}\label{lem-globint}
Let $A$ be a set, $R$ be a binary relation on $A$, $\bt \in A$, $(\con_i)_{i \in A}$ be a family of subsets of $\PPP_f(A)$, and $(\vdash_i)_{i \in A}$ be a family of relations $\vdash_i  \subseteq \con_i \times A$ such that Axioms~\ref{dn-infofr}(\ref{dn-infofr-4},\ref{dn-infofr-5},\ref{dn-infofr-7}-\ref{dn-infofr-10}) are satisfied. Then Axioms~\ref{dn-infofr}(\ref{dn-infofr-6+},\ref{dn-infofr-11}) hold if, and only if, for all $i \in A$, $X \in \con_i$ and $F \fsubset A$,
\[
X \vdash_i F \Rightarrow (\exists j \in A) (\exists Y \in \con_j) X \vdash_i j \wedge X \vdash_i Y \wedge Y \vdash_j F.
\]
\end{lem} 
\begin{pf}
For the ``only-if''-part assume that $X \vdash_i F$ and $a \in F$. By Axiom~\ref{dn-infofr}(\ref{dn-infofr-6+}) there is some $Y_a \in \con_i$ with $X \vdash_i Y_a$ and $Y_a \vdash_i a$. Set $Y = \bigcup \set{Y_a}{a \in F}$. Then $X \vdash_i Y$. Thus, $Y \in \con_i$, by Axiom~\ref{dn-infofr}(\ref{dn-infofr-4}). Because of Condition~\ref{dn-infofr}(\ref{dn-infofr-11}) there is some $j \in A$ with $X \vdash_i j$ and $Y \in \con_j$. With Lemma~\ref{lem-eq4}(\ref{lem-eq4-1}) it follows that $j R i$. Since moreover $Y \vdash_i F$, because of Requirement~\ref{dn-infofr}(\ref{dn-infofr-5}), we obtain with Axioms~\ref{dn-infofr}(\ref{dn-infofr-7},\ref{dn-infofr-10}) that $Y \vdash_j F$.

Let us next prove the ``if''-direction. We first verify Axiom~\ref{dn-infofr}(\ref{dn-infofr-6+}). Suppose that $X \in \con_i$ with $X \vdash_i a$. Then there are $e \in A$ and $Z \in \con_e$ so that $X \vdash_i e$, $X \vdash_i Z$ and $Z \vdash_e a$. With Lemma~\ref{lem-eq4}(\ref{lem-eq4-1}) it follows that $e R i$. Hence $Z \in \con_i$, by  Axiom~\ref{dn-infofr}(\ref{dn-infofr-7}). Moreover, $Z \vdash_i a$, because of Condition~\ref{dn-infofr}(\ref{dn-infofr-9}).

It remains to derive Axiom~\ref{dn-infofr}(\ref{dn-infofr-11}). Assume that $X \vdash_i Y$. Then there are $e \in A$ and $Z \in \con_e$ with $X \vdash_i e$ and $Z \vdash_e Y$, the latter implying that $Y \in \con_e$, because of  Axiom~\ref{dn-infofr}(\ref{dn-infofr-4}).
\end{pf}

As a consequence of Axioms~\ref{dn-infofr}(\ref{dn-infofr-1},\ref{dn-infofr-7},\ref{dn-infofr-8}), we have that
\begin{equation*}\label{Relim}
i R j \Longleftrightarrow \{i\} \in \con_j.
\end{equation*}
It follows that $R$ is a preorder. Moreover, it is uniquely determined by the consistency predicates and can thus be omitted from the definition of an information frame. 

Now, set
\[
\Con = \set{(i,X)}{i \in A \wedge X \in \con_i}
\]
and for $(i, X) \in \Con$ and $a \in A$,
\begin{equation}\label{eq-curr}
(i,X) \vdash a \Longleftrightarrow X \vdash_i a.
\end{equation}
Then $\vdash\subseteq \Con \times A$ and $\Con \subseteq A \times \PPP_f(A)$ such that 
\[
\Con(i) = \set{X}{(i,X)\in \Con} = \con_i.
\]

This shows that information frames can also be written in an information systems-like  style with a global entailment relation and a consistency predicate that forces consistent sets to explicitly show their consistency witness, and vice versa. 

\begin{dn}\label{dn-infsys}
Let $A$ be a set, $\bt \in A$, $\Con \subseteq A \times \PPP_f(A)$, and $\vdash \subseteq \Con \times A$. Then $\AAA = (A, \Con, \vdash, \bt)$ is an \emph{information system with witnesses} if the following conditions hold, for all $i, j, a \in A$ and all finite subsets $X, Y$ of $A$:
\begin{enumerate}
\item\label{dn-infsys-1}
$\{i\} \in \Con(i)$

\item\label{dn-infsys-2}
$Y \subseteq X \wedge X \in \Con(i) \Rightarrow Y\in \Con(i)$

\item\label{dn-infsys-3}
$(i, \emptyset) \vdash \bt$

\item\label{dn-infsys-4}
$X \in \Con(i) \wedge (i, X) \vdash Y \Rightarrow Y \in \Con(i)$

\item\label{dn-infsys-5}
$X, Y \in \Con(i) \wedge X \subseteq Y \wedge (i, X) \vdash a \Rightarrow (i, Y) \vdash a$

\item\label{dn-infsys-6}
$X \in \Con(i) \wedge (i, X) \vdash Y \wedge (i, Y) \vdash a \Rightarrow (i, X) \vdash a$

\item\label{dn-infsys-7}
$\{i\} \in \Con(j) \Rightarrow\Con(i) \subseteq \Con(j)$

\item\label{dn-infsys-8}
$\{i\}\in \Con(j) \wedge X \in \Con(i) \wedge(i,X) \vdash a \Rightarrow (j,X) \vdash a$

\item\label{dn-infsys-9}
$\{i\}\in\Con(j) \wedge X\in\Con(i) \wedge (j, X) \vdash a \Rightarrow (i,X) \vdash a$

\item\label{dn-infsys-10}
$(i,X) \vdash Y \Rightarrow (\exists (e, Z) \in \Con) (i,X)\vdash (e, Z)  \wedge (e, Z) \vdash Y$

\end{enumerate}
where $(i, X) \vdash (e, Z)$ means that $(i, X) \vdash e$ and $(i, X) \vdash Z$.
\end{dn}

Sometimes a stronger axiom than (\ref{dn-infsys-6}) is needed which reverses Axiom~(\ref{dn-infsys-10}).

\begin{lem}\label{lem-strong6}
Let $(A, \Con, \vdash, \bt)$ be an information system with witnesses. Then the following rule holds, for all $a \in A$ and $(i, X),(j, Y) \in \Con$,
\[
(i,X) \vdash (j, Y)  \wedge (j, Y) \vdash a \Rightarrow (i,X) \vdash a.
\]
\end{lem}
\begin{pf}
Since $(i,X) \vdash j$, it follows with Axiom~\ref{dn-infsys}(\ref{dn-infsys-4}) that $\{j\} \in \Con(i)$. As a consequence of Axioms~\ref{dn-infsys}(\ref{dn-infsys-7},\ref{dn-infsys-8}) we therefore have that $(i,Y) \vdash a$. Now, we can apply Axiom~\ref{dn-infsys}(\ref{dn-infsys-6}) to obtain that $(i,X) \vdash a$.
\end{pf}

As we have seen, information frames and information systems with witnesses can be derived from each other. Moreover, as will become clear next, the states associated with an information frame are the same as the states associated with the information system with witnesses generated by it, and conversely. In this sense both concepts are equivalent.

\begin{dn}\label{dn-st}
Let $\AAA = (A, \Con, \vdash, \bt)$ be an information system with witnesses. A subset $x$ of $A$ is a \emph{state} of $\AAA$ if the following three conditions hold:
\begin{enumerate}
\item\label{dn-st-1}
$(\forall F \fsubset x) (\exists i \in x) F \in \Con(i)$

\item\label{dn-st-2}
$(\forall i \in x)(\forall X \fsubset x) (\forall a \in A) [X \in \Con(i) \wedge (i,X) \vdash a \Rightarrow a \in x]$

\item\label{dn-st-3}
$(\forall a \in x) (\exists i \in x) (\exists X \fsubset x) X \in \Con(i) \wedge (i,X) \vdash a.$ 

\end{enumerate}
\end{dn}

Using the above relationship between information frames and systems this definition can be rewritten into a corresponding definition for information frames so that related frames and systems induce the same set of states. 

As follows from the definition, states are subsets of tokens that are \emph{finitely consistent} (\ref{dn-st-1}) and \emph{closed under entailment} (\ref{dn-st-2}). Furthermore, each token in a state is \emph{derivable} (\ref{dn-st-3}), i.e.\ for each token the state contains a consistent set and its witness entailing the token. Each states can therefore be understood as a theories of a model of the logic given by the information system. Condition (\ref{dn-st-3}) is a completeness requirement. 

By Condition~\ref{dn-st}(\ref{dn-st-1}) states are never empty: Choose $F$ to be the empty set. Then the state contains some $i$ with $\emptyset \in \Con(i)$.

Note that Conditions~(\ref{dn-st-1},\ref{dn-st-3}) in Definition~\ref{dn-st} can be replaced by a single requirement.

\begin{pn}\label{pn-stsing}
Let $\AAA = (A, \Con, \vdash, \bt)$ be an information system with witnesses and $x$ be a subset of $A$. Then Conditions~\ref{dn-st}(\ref{dn-st-1}) and (\ref{dn-st-3}) are equivalent to the following statement:
\begin{equation}\tag{ST}\label{st}
(\forall F \fsubset x) (\exists i \in x) (\exists X \fsubset x) X \in \Con(i) \wedge (i,X) \vdash F.
\end{equation}
\end{pn}
\begin{pf}
For the ``if''-part let $F \fsubset x$ and $a \in F$. By Condition~\ref{dn-st}(\ref{dn-st-3}) there exist $i_a \in x$ and $X_a \fsubset x$ so that $X_a \in \Con(i_a)$ and $(i_a,X_a) \vdash a$. Let $G = \set{i_a}{a \in F}$ and $X = \bigcup\set{X_a}{a \in F}$. Then $G \cup X \fsubset x$. Hence, by Condition~\ref{dn-st}(\ref{dn-st-1}), there is some $j \in x$ such that $G \cup X \in \Con(j)$. With Axiom~\ref{dn-infsys}(\ref{dn-infsys-2}) we obtain that both $X_a \in \Con(j)$ and $\{i_a\} \in \Con(j)$, for all $a \in F$, from which it follows by Axioms~\ref{dn-infsys}(\ref{dn-infsys-8}) and (\ref{dn-infsys-5}) that $(j,X) \vdash a$. Thus, $(j,X) \vdash F$.

For the ``only-if''-part we only have to show that Condition~\ref{dn-st}(\ref{dn-st-1}) holds, the other one being a special case of our assumption. Let $F \fsubset x$ again. By assumption there are $i \in x$ and $X \in \Con(i)$ with $(i,X) \vdash F$. Hence $F \in \Con(i)$, by Axiom~\ref{dn-infsys}(\ref{dn-infsys-4}).
\end{pf}

With respect to set inclusion the states of $\AAA$ form a partially ordered set, denoted by $|\AAA|$.

\begin{lem}\label{lem-infosysdom}
$|\AAA|$ is directed-complete.
\end{lem}
\begin{pf}
Let $\SSS$ be a directed subset of $|\AAA|$. It suffices to show that $\bigcup \SSS$ is a state as well:

To this end, we first verify Condition~(\ref{st}) in Proposition~\ref{pn-stsing}. Let $F \fsubset \bigcup \SSS$. Since $\SSS$ is directed, there some $x \in \SSS$ with $F \subseteq x$. Thus, there are $i \in x$ and $X \fsubset x$ with $X \in \Con(i)$ and $(i,X) \vdash F$. Since, $x \subseteq \bigcup \SSS$, we are done.

It remains to show that also Condition~\ref{dn-st}(\ref{dn-st-2}) holds. Let $i \in \bigcup\SSS$, $X \fsubset \bigcup\SSS$ and $a \in A$ such that $X \in \Con(i)$ with $(i,X) \vdash a$. Again, as $\SSS$ is directed, there is some $x \in \SSS$ with $i \in x$ and $X \subseteq x$. By Condition~\ref{dn-st}(\ref{dn-st-2}) we therefore obtain that $a \in x$. Hence, $a \in \bigcup\SSS$.
\end{pf}

As we will see next, the consistent subsets of $A$ generate a canonical basis of $|\AAA|$. For $(i,X) \in \Con$ let
\[
[X]_i = \set{a \in A}{(i,X) \vdash a}.
\]
\begin{lem}\label{lem-canba}
\begin{enumerate}
\item\label{lem-canba-1} $[X]_i$ is a state of $\AAA$, for each $(i,X) \in \Con$.
\item\label{lem-canba-2} For every $z \in |\AAA|$, the set of all $[X]_i$ with  $\{i\} \cup X \subseteq z$ is directed and $z$ is its union.
\end{enumerate}
\end{lem}
\begin{pf}
(\ref{lem-canba-1}) Conditions~(\ref{st}) and \ref{dn-st}(\ref{dn-st-2}), respectively, are immediate consequences of Axiom~\ref{dn-infsys}(\ref{dn-infsys-10}) and Lemma~\ref{lem-strong6}.

For (\ref{lem-canba-2}) let $\AAb_z = \set{[X]_i}{\{i\} \cup X \subseteq z \wedge (i,X) \in \Con}$. As $z$ is a state, there is some $j \in z$ such that $(j,\emptyset) \in \Con$. Thus, $\AAb_z$ is not empty. As a further consequence of \ref{dn-st}(\ref{dn-st-1}) we have for $(i,X), (j,Y) \in \Con$ with $\{i,j\} \cup X \cup Y \subseteq z$ that there is some $k \in z$ so that $(k, \{i,j\} \cup X \cup Y) \in \Con$. Because of \ref{dn-infsys}(\ref{dn-infsys-2}) it follows that $\{i\}, \{j\}, X, Y, X \cup Y \in \Con(k)$. With \ref{dn-infsys}(\ref{dn-infsys-8},\ref{dn-infsys-5}) we therefore obtain that $[X]_i, [Y]_j \subseteq [X \cup Y]_k$. Thus, $\AAb_z$ is directed. Obviously, $\bigcup \AAb_z \subseteq z$. Conversely, let $a \in z$. By applying \ref{dn-st}(\ref{dn-st-3}) we gain $i \in z$ and $X \fsubset z$ so that $(i,X) \vdash a$. Then $[X]_i \in \AAb_z$ and hence $a \in \bigcup \AAb_z$.
\end{pf}

This result allows characterizing the approximation relation on $A$ in terms of the entailment relation. The characterization nicely reflects the intuition that $x \ll y$ if $x$ is covered by a ``finite part'' of $y$.

\begin{pn}\label{lem-app}
For $x, y \in |\AAA|$,
\[
x \ll y \Longleftrightarrow (\exists (i,V) \in \Con) \{i\} \cup V \subseteq y \wedge (i,V) \vdash x.
\]
\end{pn}
\begin{pf}
The ``if''-part is an obvious consequence of the preceding lemma.

For the proof of the converse implication assume that $\SSS$ is a directed collection of states of $\AAA$ such that $y \subseteq \bigcup \SSS$. By the premise there is some finite subset $V$ of $y$ with consistency witness $i \in y$ such that $x \subseteq [V]_i$. It follows that $\{i\} \cup V \subseteq \bigcup \SSS$. Since $V$ is finite and $\SSS$ directed, there is some $s \in \SSS$ with $\{i\} \cup V \subseteq s$. As $s$ is a state, we obtain that also $[V]_i \subseteq s$ and hence that $x \subseteq s$. Thus, $x \ll y$.
\end{pf}

Because of Axioms~\ref{dn-infsys}(\ref{dn-infsys-1},\ref{dn-infsys-2}) we have that $\emptyset \in \Con(i)$, for all $i \in A$. Moreover, with Axioms~\ref{dn-infsys}(\ref{dn-infsys-3},\ref{dn-infsys-4}), we obtain that $\{\bt\} \in \Con(j)$, also for all $j \in A$.

\begin{lem}\label{lem-point}
\begin{enumerate}
\item\label{lem-point-1} $[\emptyset]_i = [\bt]_j$, for all $i, j \in A$.
\item\label{lem-point-2} $[\emptyset]_\bt \subseteq x$, for all $x \in |A|$.
\end{enumerate}
\end{lem}
\begin{pf}
(\ref{lem-point-1}) With \ref{dn-infsys}(\ref{dn-infsys-6},\ref{dn-infsys-3}) we have that $[\bt]_i \subseteq [\emptyset]_i$. The converse inclusion follows with \ref{dn-infsys}(\ref{dn-infsys-5}). In addition, by applying \ref{dn-infsys}(\ref{dn-infsys-8},\ref{dn-infsys-9}), we gain that $[\emptyset]_\bt = [\emptyset]_i$. The statement is now an easy consequence.

(\ref{lem-point-2}) As states are nonempty, there is some $i \in x$. Moreover, by Axiom~\ref{dn-infsys}(\ref{dn-infsys-3}), $(i,\emptyset) \vdash \bt$. Thus, $\bt \in x$, because of \ref{dn-st}(\ref{dn-st-2}). Hence, by applying the same rule again, we obtain that $[\emptyset]_\bt \subseteq x$.
\end{pf} 

\begin{lem}\label{lem-relsup}
Let $x, y, z \in |\AAA|$ so that $x, y \subseteq z$. Then 
\[
\bigcup\set{[Z]_k}{(k, Z) \in \Con\mbox{} \wedge k \in z \wedge Z \fsubset x \cup y}
\]
is the least upper bound of $x$ and $y$ in $\low z$.
\end{lem}
\begin{pf}
As in the proof of Lemma~\ref{lem-canba} it follows that $\set{[Z]_k}{k \in z \wedge Z \fsubset x \cup y}$ is directed. Thus, $\bigcup\set{[Z]_k}{k \in z \wedge Z \fsubset x \cup y} \in |\AAA|$. 

Let $(i, X) \in \Con$ such that $\{ i \} \cup X \subseteq x$. Then $i \in z$ and $X \subseteq x \cup y$. Thus, $[X]_i \in \set{[Z]_k}{k \in z \wedge Z \fsubset x \cup y}$. With Lemma~\ref{lem-canba}(\ref{lem-canba-2}) it follows that $x \subseteq \bigcup \{\, [Z]_i \mid i \in z \wedge Z \fsubset x \cup y \,\}$. In the same way we obtain that $y \subseteq \bigcup \set{[Z]_i}{i \in z \wedge Z \fsubset x \cup y}$.

Finally, let $u \in |\AAA|$ with $x, y \subseteq u \subseteq z$. Moreover, let $(k, Z) \in \Con$ so that $k \in z$ and $Z \fsubset x \cup y$. Then $Z \fsubset u$. Hence, there exist $e \in u$ such that $Z \in \Con(e)$. Since $\{ e, k \} \subseteq z$, it follows with  Property~(\ref{st})  that there is some $(d, U) \in \Con$ with $\{ d \} \cup U \subseteq z$ and $(d, U) \vdash \{ e, k \}$. With \ref{dn-infsys}(\ref{dn-infsys-3},\ref{dn-infsys-7}-\ref{dn-infsys-9}) we now obtain that $Z \in \Con(d)$ and $[Z]_k = [Z]_d = [Z]_e$. Thus, $[Z]_k \subseteq	u$. 
\end{pf}

Let us now sum up what we have shown so far.

\begin{thm}\label{tm-ldom}
Let $\AAA = (A, \Con, \vdash, \bt)$ be an information system with witnesses. Then $\LLL(\AAA) = (|\AAA|, \subseteq, [\emptyset]_\bt)$ is an L-domain with basis $\widehat{\Con} = \set{[X]_i}{ (i,X) \in \Con}.$
\end{thm}

Next, we study when $\LLL(\AAA)$ is algebraic. As a consequence of Lemma~\ref{lem-strong6} and Proposition~\ref{lem-app} we obtain:

\begin{lem}\label{lem-comp}
For $(i,Z) \in \Con$ the following two statements are equivalent:
\begin{enumerate}
\item\label{lem-comp-1} $[Z]_i \ll [Z]_i$
\item\label{lem-comp-2} $(\exists (j, V) \in \Con)  (i, Z) \vdash (j, V) \wedge (j,V) \vdash (j, V) \wedge (j,V) \vdash [Z]_i$.
\end{enumerate}
\end{lem}

\begin{dn}\label{dn-refl}
Let $(A, \Con, \vdash, \bt)$ be an information system with witnesses. An element $(j,V) \in \Con$ is called \emph{reflexive} if $(j,V) \vdash (j, V)$.
\end{dn}

Let $\Con_\mathrm{refl}$ denote the subset of reflexive elements of $\Con$. Obviously, $[V]_j$ is compact, for every  $(j,V) \in \Con_\mathrm{refl}$.

\begin{lem}\label{lem-combas}
The following two statements are equivalent:
\begin{enumerate}
\item\label{lem-combas-1}
For every $z \in |\AAA|$, the set of all $[V]_j$ with $(j, V) \in \Con_\mathrm{refl}$  and $\{ j \} \cup V \subseteq z$ is directed and its union is $z$.

\item\label{lem-combas-2}
The information system $\AAA$ satisfies Condition~(\ref{alg}) saying that for all $(i,X) \in \Con$ and $F \fsubset A$, 
\begin{equation}\tag{ALG}\label{alg}
(i, X) \vdash F  \Rightarrow (\exists (j,V) \in \Con_\mathrm{refl}) (i, X) \vdash  (j, V) \wedge (j, V) \vdash F.
\end{equation}
\end{enumerate}
\end{lem}
\begin{pf}
The ``if''-part is obvious. For the ``only if''-part let $$\BB_z = \set{[V]_j}{(j, V) \in \Con_\mathrm{refl}\mbox{} \wedge \{ j \} \cup V \subseteq z}.$$ Then it follows as in the proof of Lemma~\ref{lem-canba}(\ref{lem-canba-2}) that $\BB_z$ is not empty. Let $(i, U), (j, V) \in \Con_\mathrm{refl}$ with $\{ i, j \} \cup U \cup V \subseteq z$. Then it follows with Property~(\ref{st}) that there is some $(k, X) \in \Con$ with $\{ k \} \cup X \subseteq z$ such that $(k, X) \vdash \{ i, j \} \cup U \cup V$. By Condition~(\ref{alg}), we furthermore obtain some $(e, Z)  \in \Con_\mathrm{refl}$ so that $(k, X) \vdash (e, Z)$ and $(e, Z) \vdash \{ i, j \} \cup U \cup V$. Thus,  $[U]_i, [V]_j \subseteq [Z]_e \subseteq [X]_k \subseteq z$, which shows that $\BB_z$ is directed. 

It remains to show that $z \subseteq \bigcup \BB_z$. Let to this end $a \in z$. Because of  \ref{dn-st}(\ref{dn-st-3}) there is some $(k, X) \in \Con$ with $\{ k \} \cup X \subseteq z$ and $(k, X) \vdash a$. As we have just seen, by Condition~(\ref{alg}) there exists $(j, Y) \in \Con_\mathrm{refl}$ such that $\{ j \} \cup Y \in  [X]_k \subseteq z$ and $a \in [Z]_j$, which means that $a \in \bigcup\BB_z$.
\end{pf}

According to Lemma~\ref{lem-globint}, in the presence of Axioms~\ref{dn-infsys}(\ref{dn-infsys-4},\ref{dn-infsys-5},\ref{dn-infsys-7},\ref{dn-infsys-8},\ref{dn-infsys-9}) Condition~\ref{dn-infsys}(\ref{dn-infsys-10}) holds, exactly if for all $(i, X) \in \Con$, $a \in A$ and $F \fsubset A$ the following two requirements hold:
\begin{gather}
(i, X) \vdash a \Rightarrow (\exists Z \in \Con(i)) (i, X) \vdash Z \wedge (i, Z) \vdash a \label{10-1} \\
(i, X) \vdash F \Rightarrow (\exists j \in A) (i, X) \vdash j \wedge F \in \Con(j). \label{10-2}
\end{gather}

This allows to simplify Condition~(\ref{alg}).

\begin{lem}\label{lem-salg}
Condition~(\ref{alg}) holds if, and only if, the following Condition~(\ref{salg}) is satisfied for all $(i, X) \in \Con$ and $a \in A$,
\begin{equation} \tag{SALG}\label{salg}
(i, X) \vdash a \Rightarrow (\exists Z \in \Con(i)) (i, X) \vdash Z \wedge (i, Z) \vdash Z \wedge (i, Z) \vdash a.
\end{equation}
\end{lem}
\begin{pf}
Assume first that (\ref{alg}) holds, and let $(i, X) \in \Con$ and $a \in A$ with $(i, X) \vdash a$. Then there is some reflexive $(j, Z) \in \Con$ such that $(i, X) \vdash (j, Z)$ and $(j, Z) \vdash a$. In particular, we obtain that $\{ j \} \in \Con(i)$ and hence that $(i, Z) \vdash Z$ as well as $(i, Z)  \vdash a$.

Next, suppose that (\ref{salg}) is satisfied, and let $(i, X) \in \Con$ and $F \fsubset A$ with $(i, X) \vdash F$. Then, for every $a \in F$, there is some $Z_a \in \Con(i)$ so that $(i, X) \vdash Z_a$, $(i, Z_a) \vdash Z_a$ and $(i, Z_a) \vdash a$. For $Z = \bigcup_{a \in F} Z_a$ it follows that $(i, X) \vdash Z$, from which we obtain that $Z \in \Con(i)$. Moreover, we have that $(i, Z) \vdash Z$ and $(i, Z) \vdash F$. Because of Condition~(\ref{10-2}) there is now some $j \in A$ with $(i, Z) \vdash j$ and $Z \in \Con(j)$. With Axioms~\ref{dn-infsys}(\ref{dn-infsys-6}) and \ref{dn-infsys}(\ref{dn-infsys-9}) it follows that $(i, X) \vdash j$, $(j, Z) \vdash (j, Z)$ and $(j, Z) \vdash F$, as was to be shown.
\end{pf}

\begin{thm}\label{tm-alg}
Let $\AAA = (A, \Con, \vdash, \bt)$ be an information system with witnesses. Then $\LLL(\AAA)$ is algebraic if, and only if,  information system $\AAA$ satisfies Condition~(\ref{alg}).
\end{thm}

As well, we will provide a condition which guaranties that $\LLL(\AAA)$ is a bounded-complete. 

\begin{thm}\label{tm-bc}
Let $\AAA = (A, \Con, \vdash, \bt)$ be an information system with witnesses. Then $\LLL(\AAA)$ is bounded-complete, and hence a bc-domain, if  information system $\AAA$ satisfies Condition~(\ref{bc}) saying that for all $X \fsubset A$ and $i, j \in A$,
\begin{equation}\tag{BC}\label{bc}
(i, X), (j, X) \in \Con \Rightarrow (\forall a \in A)[(i, X) \vdash a \Leftrightarrow (j, X) \vdash a].
\end{equation}
\end{thm}
\begin{pf}
Since $|\AAA|$ is directed-complete, it suffices to show that any pair of elements that is bounded above has a least upper bound. Let to this end $x, y \subseteq z, z'$. We will prove that $x \sqcup^z y = x \sqcup^{z'} y$.

Let $a \in x \sqcup^z y$. Then there are $(k, Z) \in \Con$ such that $k \in z$, $Z \subseteq x \cup y$, and $a \in [Z]_k$. It follows that $Z \fsubset z'$. Thus, because of \ref{dn-st}(\ref{dn-st-1}), $Z \in \Con(k')$, for some $k' \in z'$. As $[Z]_k = [Z]_{k'}$, by Condition~(\ref{bc}), we obtain that $a \in x \sqcup^{z'} y$.   
\end{pf}

It is unknown whether the requirement on $\AAA$ is also necessary. In what follows, however, we will conversely show that every L-domain $\DDD$ defines an information system with witnesses the associated L-domain of which is isomorphic to $D$. In case that $\DDD$ is bounded-complete this information system will satisfy Condition~(\ref{bc}).

Let $\DDD = (D, \sqsubseteq)$ be an L-domain with basis $B$ and least element $\bot$. Set
\[
\III(\DDD) = (B, \Con, \vdash, \bot)
\]
with
\[
 \Con = \set{(i, X)}{i \in B \wedge X \fsubset \low i \cap B} 
 \]
and
\[
 (i, X) \vdash a \Longleftrightarrow a \ll \bigsqcup\nolimits^i X.
\]

\begin{lem}\label{pn-dominf}
$\III(\DDD)$ is an information system with witnesses. 
\end{lem}
\begin{pf}
All conditions in Definition~\ref{dn-infsys} are easy consequences of Lemma~\ref{lem-preordprop}. In particular, Condition~\ref{dn-infsys-10} follows from the interpolation law.
\end{pf}

\begin{lem}\label{lem-stdir}
Every state of $\III(\DDD)$ is a directed subset of $D$.
\end{lem}
\begin{pf}
As we have already seen, states are not empty. Let $x \in |\III(\DDD)|$ and $a, b \in x$. Then it follows with \ref{dn-st}(\ref{dn-st-1}) that $\{ a, b \} \in \Con(i)$, for some $i \in x$. Thus $a, b \sqsubseteq i$.
\end{pf}

It follows that $\bigsqcup x$ exists in $D$, for every $x \in |\III(D)|$. For $x \in |\III(D)|$ set 
\[
\sp\nolimits_\DDD(x) = \bigsqcup x.
\]
Then $\fun{\sp_\DDD}{|\III(\DDD)|}{\DDD}$ is Scott continuous.

\begin{lem}\label{lem-domst}
For $\alpha \in D$, $\set{a \in B}{a \ll \alpha}$ is a state of $\III(\DDD)$.  
\end{lem}
\begin{pf}
Condition~(\ref{st}) is an immediate consequence of the interpolation law and Condition~\ref{dn-st}(\ref{dn-st-2}) is obvious.
\end{pf}

Set 
\[
\st\nolimits_\DDD(\alpha) = \set{a \in B}{a \ll \alpha},
\]
for $\alpha \in D$. Then $\fun{\st_\DDD}{\DDD}{|\III(\DDD)|}$ is Scott continuous as well. Since $B$ is a basis of $\DDD$, we have that $\sp_\DDD(\st_\DDD(\alpha)) = \alpha$.

\begin{lem}\label{lem-invstsp}
For $x \in |\III(\DDD)|$, $\st_\DDD(\sp_D(x)) = x$.
\end{lem}
\begin{pf}
We have that
\[
\st\nolimits_\DDD(\sp\nolimits_\DDD(x)) = \set{a \in B}{a \ll \bigsqcup x} = \set{a \in B}{(\exists b \in x) a \ll b}.
\]

If $a \ll b$, for some $b \in x$, we obtain by \ref{dn-st}(\ref{dn-st-3}) that there is some $(i, X) \in \Con$ with $\{ i \} \cup X \subseteq x$ such that $a \ll b \ll \bigsqcup^i X$, from which it follows that $a \ll \bigsqcup^i X$. Hence, $(i, X) \vdash a$. By \ref{dn-st}(\ref{dn-st-2}) we gain that $a \in x$.

Conversely, if $a \in x$, then, again by \ref{dn-st}(\ref{dn-st-3}), there is some $(i, X) \in \Con$ so that $\{ i \} \cup X \subseteq x$ and $(i, X) \vdash a$. It follows that $a \ll \bigsqcup^i X \sqsubseteq i$, and hence that $a \ll i$.
\end{pf}

Thus, both functions are inverse to each other, which shows that $D$ is isomorphic to $|\III(D)|$.

\begin{thm}\label{tm-dominsys}
Let $\DDD$ be an L-domain. Then $\III(\DDD)$ is an information system with witnesses such that $\DDD$ and $\LLL(\III(\DDD))$ are isomorphic. In addition,
\begin{enumerate}
\item\label{tm-dominsys-1}
$\DDD$ is algebraic if, and only if, the information system $\III(\DDD)$ satisfies Condition~(\ref{alg}).
\item\label{tm-dominsys-2}
$\DDD$ is bounded-complete if, and only if, Condition~(\ref{bc}) holds in $\III(\DDD)$. 
\end{enumerate}
\end{thm}
\begin{pf}
It remains to demonstrate Statements~(\ref{tm-dominsys-1}) and (\ref{tm-dominsys-2}). Because of Theorems~\ref{tm-alg} and \ref{tm-bc}, and as $\DDD$ and $\LLL(\III(\DDD))$ are isomorphic, it suffices  to consider just the ``only if''-parts, which are obvious, however.
\end{pf}

In the remainder of this section we consider two special cases.

\begin{pn}\label{pn-op}
Let $\TTT = (\{ \bt \}, \Con_\TTT, \vdash_\TTT, \bt)$, where
\[
\Con_\TTT = \{(\bt, \emptyset), (\bt, \{\bt\})\} \text{ and } \vdash_\TTT = \Con_\TTT \times \{\bt\}.
\]
Then $\TTT$ is an information system with witnesses satisfying Conditions~(\ref{alg}) and (\ref{bc}). $|\TTT|$ is the one-point domain.
\end{pn}

Let $\AAA_{1} = (A_1, \Con_1, \vdash_1, \bt_1)$ and $\AAA_{2} = (A_2, \Con_2, \vdash_2, \bt_2)$ be information systems with witnesses, and $\pr_1$ and $\pr_2$, respectively, be the canonical projections of $A_1 \times A_2$ onto the first and second component. Set $A_\times = A_1 \times A_2$, $\bt_\times = (\bt_1, \bt_2)$,
\[
\Con_\times = \set{((i,j), X) \in A_\times \times \PPP_f(A_\times)}{\pr\nolimits_1(X) \in \Con_1(i) \wedge \pr\nolimits_2(X) \in \Con_2(j)}, 
\]
and for $((i,j), X) \in \Con_\times$ and $(a_1, a_2) \in A_\times$ define
\[
((i,j), X) \vdash_\times (a_1, a_2) \Longleftrightarrow (i, \pr\nolimits_1(X)) \vdash_1 a_1 \wedge (j, \pr\nolimits_2(X)) \vdash_2 a_2.
\]
Then $\AAA_{\times} = (A_\times, \Con_\times, \vdash_\times, \bt_\times)$ is an information system with witnesses, the \emph{product} of $\AAA_{1}$ and $\AAA_{2}$.

\begin{lem}\label{lem-prod}
For $z \in |\AAA_\times|$ and $\nu = 1, 2$, the following two statements hold:
\begin{enumerate}
\item\label{lem-prod-1}
$\pr\nolimits_\nu (z) \in |\AAA_\nu|$.

\item\label{lem-prod-2}
$z = \pr\nolimits_1(z) \times \pr\nolimits_2(z)$.
\end{enumerate}
\end{lem}
\begin{pf}
(\ref{lem-prod-1}) Without restriction let $\nu = 1$. We only verify Condition~\ref{dn-st}(\ref{dn-st-2}), the other two being obvious. Let $a_1 \in A_1$ and $(i_1, Y_1) \in \Con_1$ with $\{ i_1 \} \cup Y_1 \subseteq \pr_1(z)$ and $(i_1, Y_1) \vdash_1 a_1$. Then there are $i_2 \in A_2$ and $X \fsubset A_\times$ with $\pr_1(X) = Y_1$ and $\{ (i_1, i_2)\} \cup X \subseteq z$. Let $a_2$ be some element of $\pr_2(z)$. Hence, by~\ref{dn-st}(\ref{dn-st-3}), there is some $((j_1, j_2), Z) \in \Con_\times$ with $\{ (j_1, j_2)\} \cup Z \subseteq z$ and $(j_2, \pr_2(Z) ) \vdash_2 a_2$. Now, by applying Condition~(\ref{st}), we obtain some $((k_1, k_2), V) \in \Con_\times$ with $\{ (k_1, k_2)\} \cup V \subseteq z$ so that $((k_1, k_2), V) \vdash_\times \{ (i_1, i_2), (j_1, j_2)\} \cup X \cup Z $. It follows that $((k_1, k_2), V) \vdash_\times (a_1, a_2)$. Thus, $(a_1, a_2) \in z$, by \ref{dn-st}(\ref{dn-st-2}), which means that $a_1 \in \pr_1(z)$.

(\ref{lem-prod-2}) is easily shown by applying Conditions~\ref{dn-st}(\ref{dn-st-2}, \ref{dn-st-3}).
\end{pf}

\begin{pn}\label{pn-prodprop}
Let $\AAA_{1}$ and $\AAA_{2}$ be information systems with witnesses. Then $\AAA_{\times}$ too is an information system with witnesses and the L-domains $|\AAA_\times|$ and $|\AAA_1| \times |\AAA_2|$ are isomorphic. Moreover,
\begin{enumerate}
\item\label{pn-prodprop-1}
If both $\AAA_1$ and $\AAA_2$ satisfy Condition~(\ref{alg}), so does $\AAA_\times$.

\item\label{pn-prodprop-2}
If both $\AAA_1$ and $\AAA_2$ satisfy Condition~(\ref{bc}), so does $\AAA_\times$.
\end{enumerate}
\end{pn}

\section{Approximable mappings}\label{sec-appmp}

In the next step we want to turn the collection of information systems with witnesses into a category.  The appropriate morphisms are relations similar to entailment relations. In the case of information frames one has to consider families of such relations.

\begin{dn}\label{dn-am}\sloppy
An \emph{approximable mapping} $H$ between information systems with witnesses $\AAA = (A, \Con, \vdash, \bt)$ and $\AAA' = (A', \Con', \vdash', \bt')$, written $\apmap{H}{\AAA}{\AAA'}$, is a relation between $\Con$ and $A'$ satisfying the following five conditions, for all $i, j \in A$, $X, X' \fsubset A$, $k \in A'$ and $Y, F \fsubset A'$ with $X \in \Con(i)$ and $Y \in \Con'(k)$:
\begin{enumerate}
\item\label{dn-am-1}
$(i, X) H (k, Y) \wedge (k, Y) \vdash' b \Rightarrow (i, X) H b$

\item\label{dn-am-2}
$X' \in \Con(i) \wedge X \subseteq X' \wedge (i, X) H b \Rightarrow (i, X') H b$

\item\label{dn-am-3}
$(i, X) \vdash X' \wedge (i, X') H b \Rightarrow (i, X) H b$

\item\label{dn-am-4}
$\{ i \} \in \Con(j)  \wedge (i, X) H b \Rightarrow (j, X) H b$

\item\label{dn-am-5}
$(i, X) H F \Rightarrow (\exists (c, U) \in \Con) (\exists (e, V) \in \Con')  (i, X) \vdash (c, U) \wedge (c, U) H (e, V) \wedge (e, V) \vdash' F$ 

\item\label{dn-am-6}
$(\bt, \emptyset) H \bt'$.
\end{enumerate}
Here, $(i, X) H Y$ means that $(i, X) H c$, for all $c \in Y$,  and $(i, X) H (k, Y)$ that $(i, X) H k$ as well as $(i, X) H Y$.
\end{dn}

In applications it is sometimes preferable to have Condition~(\ref{dn-am-5}) split up into two conditions which state interpolation for the domain and the range of the approximable mapping, separately.

\begin{lem}\label{pn-amint}
Let $(A, \Con, \vdash, \bt)$ and $(A', \Con', \vdash', \bt')$ be information systems with witnesses. Then, for any $H \subseteq \Con \times A'$, $(i, X) \in \Con$, and $F \fsubset A'$, Condition~\ref{dn-am}(\ref{dn-am-5}) is equivalent to the following Conditions~(\ref{pn-amint-1}) and (\ref{pn-amint-2}):
\begin{enumerate}
\item\label{pn-amint-1}
$(i, X) H F \Rightarrow (\exists (c, U) \in \Con) (i, X) \vdash (c, U) \wedge (c, U) H F$

\item\label{pn-amint-2}
$(i, X) H F \Rightarrow (\exists (e, V) \in \Con') (i, X) H (e, V) \wedge (e, V) \vdash' F$.
\end{enumerate}
\end{lem}
\begin{pf}
The ``if''-part follows with \ref{dn-infsys}(\ref{dn-infsys-4}) as well as \ref{dn-am}(\ref{dn-am-4},\ref{dn-am-3}), and \ref{dn-am}(\ref{dn-am-1}), respectively. The ``only if''-part is obvious.
\end{pf}

Similar to Lemma~\ref{lem-strong6} a strengthening of Axiom~\ref{dn-am}(\ref{dn-am-3}) can be derived. It reverses the implication in Lemma~\ref{pn-amint}(\ref{pn-amint-1}).

\begin{lem}\label{lem-amstrong3}
Let $H$ be an approximable mapping between information systems $\AAA$ and $\AAA'$ with witnesses. Then for all $(i, X), (j, Y) \in \Con$ and $b \in A'$,
\[
(i, X) \vdash (j, Y) \wedge (j, Y) H b \Rightarrow (i, X) H b.
\]
\end{lem}

As has already been mentioned, entailment relations are special approximable mappings. For $(i, X) \in \Con$ and $a \in A$, set $(i, X) \Id_\AAA a$ if $(i, X) \vdash a$. Then $\apmap{\Id_{\AAA}}{\AAA}{\AAA}$ such that for all $\apmap{H}{\AAA}{\AAA'}$, $H \circ \Id_{\AAA'} = H = \Id_\AAA \circ H$, where for approximable mappings $\apmap{H}{\AAA}{\AAA'}$ and $\apmap{G}{\AAA'}{\AAA''}$ their composition $\apmap{H \circ G}{\AAA}{\AAA''}$ is defined by
\[
(i, X) (H \circ G) c \Longleftrightarrow (\exists (j, Y) \in \Con') (i, X) H (j, Y) \wedge (j, Y) G c.
\]

Let $\mathbf{ISW}$ be the category of information systems with witnesses and approximable mappings and $\mathbf{aISW}$, $\mathbf{bcISW}$, and $\mathbf{abcISW}$, respectively, be the full subcategories of information systems with witnesses that satisfy Condition~(\ref{alg}), Condition~(\ref{bc}), or both of them.

\begin{pn}\label{pn-term}
The one-point information system $\TTT$ with witnesses  is a terminal object in $\mathbf{ISW}$.
\end{pn}
\begin{pf}
\sloppy Let $\AAA' = (A', \Con', \vdash', \bt')$ be an information system with witnesses and $H = \Con' \times \{ \bt \}$. It suffices to show that $\apmap{H}{\AAA'}{\TTT}$. We only verify Condition~\ref{dn-am}(\ref{dn-am-5}), the others being obvious.

Let to this end $(i, X) \in \Con'$ with $(i, X) H F$, where $F = \emptyset$ or $F = \{ \bt \}$. Then $(i, X) \vdash' (\bt', \emptyset)$, $(\bt', \emptyset) H (\bt, \emptyset)$, and $(\bt, \emptyset) \vdash_T F$.
\end{pf}

As $\TTT$ satisfies both (\ref{alg}) and (\ref{bc}), it is of course also terminal in $\mathbf{aISW}$, $\mathbf{bcISW}$ and $\mathbf{abcISW}$. 

For two information systems  $\AAA_{1} = (A_1, \Con_1, \vdash_1, \bt_1)$ and $\AAA_{2} = (A_2, \Con_2, \vdash_2, \bt_2)$ with witnesses define the relations $\Pr_\nu \subseteq \Con_\times \times A_\nu$, for $\nu = 1, 2$, by
\[
((i_1, i_2), X) \Pr\nolimits_\nu a_\nu \Longleftrightarrow (i_\nu, \pr_\nu(X)) \vdash_\nu a_\nu.
\]

\begin{lem}\label{lem-proj}
For $\nu = 1, 2$, $\apmap{\Pr_\nu}{\AAA_\times}{\AAA\nu}$.
\end{lem}
\begin{pf}
Again, we verify only Condition~\ref{dn-am}(\ref{dn-am-5}). Let $((i_1, i_2), X) \in \Con_\times$ and $F \fsubset  A_\nu$ with $((i_1, i_2), X) \Pr_\nu F$. Then $(i_\nu, \pr_\nu(X)) \vdash_\nu F$. Hence, there are $(j_\nu, Y_\nu) , (k_\nu, Z_\nu) \in \Con_\nu$ so that $(i_\nu, \pr_\nu(X)) \vdash_\nu (j_\nu, Y_\nu) \vdash_\nu (k_\nu, Z_\nu) \vdash_\nu F$. Set $Y = Y_1 \times \{ \bt_2 \}$ and $j_2 = \bt_2$, if $\nu = 1$, and $Y = \{ \bt_1 \} \times Y_2$ as well as $j_1 = \bt_1$, otherwise. It follows that $((i_1, i_2), X) \vdash_\times ((j_1, j_2), Y) \Pr_\nu (k_\nu, Z_\nu) \vdash_\nu F$.
\end{pf}

\begin{pn}\label{pn-infcatprod}
For information systems $\AAA_1$ and $\AAA_2$ with witnesses, $(\AAA_\times, \Pr_1, \Pr_2)$ is their categorical product.
\end{pn}

Note that for approximable mappings $\apmap{H_1}{\AAA}{\AAA_1}$ and $\apmap{H_2}{\AAA}{\AAA_2}$ the mediating morphism $\apmap{\pair{H_1, H_2}}{\AAA}{\AAA_\times}$ is given by
\[
(i, X) \pair{H_1, H_2} (a_1, a_2) \Longleftrightarrow (i, X) H_1 a_1 \wedge (i, X) H_2 a_2.
\]

As we have already seen, there is a close connection between information systems with witnesses and L-domains. It can be extended to the corresponding morphisms, i.e.\ approximable mappings and Scott continuous functions, so that we obtain an equivalence between $\mathbf{ISW}$ and $\mathbf{L}$.

Let $\AAA = (A, \Con, \vdash, \bt)$ and $\AAA' = (A', \Con', \vdash', \bt')$ be information systems with witnesses and $\apmap{H}{\AAA}{\AAA'}$.

\begin{lem}\label{lem-apsc}
For  $x \in |\AAA|$, 
\[
\set{a \in A'}{(\exists (i, X) \in \Con) \{ i \} \cup X \subseteq x \wedge (i, X) H a} \in |\AAA'|.
\]
\end{lem}
\begin{pf}
For the verification of Condition~\ref{dn-st}(\ref{dn-st-2}) set 
\[
y = \{\, a \in A' \mid (\exists (i, X) \in \Con) \{ i \} \cup X \subseteq x \wedge (i, X) H a \,\}
\] 
and let $a \in A'$. Moreover, let $(j, Y) \in \Con'$ such that $\{ j \} \cup Y \subseteq y$ and $(j, Y) \vdash' a$. Then there exists $(i, X) \in \Con$ with $\{ i \} \cup X \subseteq x$ and $(i, X) H j$. In addition, for every $e \in Y$, there is $(b_e, Z_e) \in \Con$ so that $\{ b_e \} \cup Z_e \subseteq x$ and $(b_e, Z_e) H e$. Set $F = \{ i\} \cup X \cup \bigcup \set{Z_e}{e \in Y} \cup \bigcup \set{b_e}{e \in Y}$. By (\ref{st}) there is thus some $(k, U) \in \Con$ with $\{ k \} \cup U \subseteq x$ and $(k, U) \vdash F$. It follows that 
\[
(k, U ) \vdash (i, X) \text{ and } (i, X) H j,
\]
whence, by Lemma~\ref{lem-amstrong3}, we obtain that $(k, U) H j$. Similarly, for every $e \in Y$, we have $(k, U) \vdash (b_e, Z_e)$ and $(b_e, Z_e) H e$, and hence that $(k, U) H e$. So, we gain that $(k, U) H (j, Y)$. Since $(j, Y) \vdash' a$, an application of \ref{dn-am}(\ref{dn-am-1}) yields $(k, U) H a$. Thus, $a \in y$.

It remains to verify Condition~(\ref{st}).  Let to this end, $F \fsubset y$.  Then, for any $e \in F$, there is some $(i_e, X_e) \in \Con$ with $\{ i_e \} \cup X_e \subseteq x$ and $(i_e, X_e) H e$. Set $K = \set{i_e}{e \in F} \cup \bigcup \set{X_e}{e \in F}$. Then $K \fsubset x$. Thus, it follows with Condition~(\ref{st}) that there is $(j, U) \in \Con$ so that $\{ j \} \cup U \subseteq x$ and $(j, U) \vdash K$. In particular, we have that $(j, U) \vdash (i_e,  X_e)$, for every $e \in F$. With Lemma~\ref{lem-amstrong3} we therefore obtain that $(j, U) H F$.  Because of Lemma~\ref{pn-amint}(\ref{pn-amint-2}) there is now some $(k, V) \in \Con'$ with $(j, U) H (k, V)$ and $(k, V) \vdash F$. It remains to show that  $\{ k \} \cup V \subseteq y$, which, however, is a consequence of $(j, U) H (k, V)$.
 \end{pf}

This allows us to define a function $\fun{\LLL(H)}{\LLL(\AAA)}{\LLL(\AAA')}$ by
\[
\LLL(H)(x) =  \{\, a \in A' \mid (\exists (i, X) \in \Con) \{ i \} \cup X \subseteq x \wedge (i, X) H a \,\}.
\]

\begin{lem}\label{lem-apmapst}
$\LLL(H)$ is Scott continuous.
\end{lem}
\begin{pf}
Obviously, $\LLL(H)$ is monotone. Let $\SSS$ be a directed subset of $|\AAA|$. Then it remains to show that $\LLL(H)(\bigcup \SSS) \subseteq \bigcup \LLL(H)(\SSS)$, the converse inclusion being a consequence of monotonicity.

Let $b \in \LLL(H)(\bigcup \SSS)$. Then there is some $(i, X) \in \Con$ with $\{ i \} \cup X \subseteq \bigcup \SSS$ and $(i, X) H b$. Since $\SSS$ is directed and $X$ finite, it follows that $\{ i \} \cup X \subseteq x$, for some $x \in \SSS$. Thus, $b \in \LLL(H)(x)$.
\end{pf}

\begin{lem}\label{lem-funcl}
$\fun{\LLL}{\mathbf{IWS}}{\mathbf{L}}$ is a functor.
\end{lem}
\begin{pf}
Because of Conditions~\ref{dn-st}(\ref{dn-st-2},\ref{dn-st-3}) we have that 
\[
\LLL(\Id\nolimits_\AAA)(x) = \set{a \in A}{(\exists (i, X) \in \Con) \{ i \} \cup X \subseteq x \wedge (i, X) \vdash a} = x.
\]
Thus, $\LLL(\Id_\AAA) = \id_{|\AAA|}$, where $\id_{|\AAA|}$ is the identity function on $|\AAA|$. It remains to show functoriality.

Let to this end $\AAA'' = (A'', \Con'', \vdash'', \bt'')$ be a further information system with witnesses, and $\apmap{H}{\AAA}{\AAA'}$ as well as $\apmap{G}{\AAA'}{\AAA''}$. Moreover, let $c \in \LLL(G)(\LLL(H)(x))$, for $x \in |\AAA|$. Then there is some $(j, Y) \in \Con'$ with $\{ j \} \cup Y \subseteq \LLL(H)(x)$ and $(j, Y) G c$. Let $F = \{ j \} \cup Y$. Then it follows that for every $b \in F$ there is some $(i_b, X_b) \in \Con$ so that $\{ i_b \} \cup X_b \subseteq x$ and $(i_b, X_b) H b$. Set $K = \set{i_b}{b \in F} \cup \bigcup\set{X_b}{b \in F}$. Then $K \fsubset x$. Hence, there is some $(k, Z) \in \Con$ with $\{ k \} \cup Z \subseteq x$ and $(k, Z) \vdash K$. Consequently, $(k, Z) H (j, Y)$. This shows that $c \in \LLL(H \circ G)(x)$. 

Now, conversely, let $c \in \LLL(H \circ G)(x)$. Then there is some $(i, X) \in \Con$ so that $\{ i \} \cup X \subseteq x$ and $(i, X) (H \circ G) c$. It follows that there is also some $(j, Y) \in \Con'$ with $(i, X) H (j, Y)$ and $(j, Y) G c$. Thus, $\{ j \} \cup Y \subseteq \LLL(H)(x)$. So, we have that $c \in \LLL(G)(\LLL(H)(x))$.
\end{pf}

Let us next consider the converse situation in which we went from L-domains to information systems with witnesses. As we will see, every Scott continuous function $\fun{f}{D}{D'}$ between L-domains $\DDD$ and $\DDD'$ defines an approximable mapping $\apmap{\III(f)}{\III(\DDD)}{\III(\DDD')}$. 

Let $\DDD$ and $\DDD'$, respectively, have bases $B$ and $B'$. Then, for $i \in B$, $X \fsubset \low i \cap B$, and $a \in B'$, set
\[
(i, X) \III(f) a \Longleftrightarrow a \ll' f(\bigsqcup\nolimits^i X).
\]

\begin{lem}\label{lem-scam}
$\apmap{\III(f)}{\III(\DDD)}{\III(\DDD')}$. 
\end{lem}
\begin{pf}
Let $i \in B$, $X \fsubset \low i \cap B$, $k, b \in B'$, and $Y, S \fsubset \low k \cap B'$. We have to verify Conditions~\ref{dn-am}(\ref{dn-am-1}-\ref{dn-am-6}).

(\ref{dn-am-1})  Assume that $(i, X) \III(f) (k, Y)$ and $(k, Y) \vdash' b$. Then we have that $b \ll' \bigsqcup\nolimits^k Y \ll' f(\bigsqcup\nolimits^i X)$, where in the last case we had to apply the interpolation law first. It follows that $b \ll' f(\bigsqcup\nolimits^i X)$. Thus,  $(i, X) \III(f) b$.

Condition~(\ref{dn-am-3}) follows in a similar way, and Conditions~(\ref{dn-am-2}), (\ref{dn-am-4}) and (\ref{dn-am-6}) are obvious. We consider only Condition~(\ref{dn-am-5}).

Assume that $(i, X) \III(f) S$. Then $S \ll' f(\bigsqcup\nolimits^i X)$. Because of the interpolation law there is some $e \in B'$ with $S \ll' e \ll' f(\bigsqcup\nolimits^i X)$.  As $f$ is Scott continuous, we have that $f(\bigsqcup\nolimits^i X) = \bigsqcup f(\set{a \in B}{a \ll \bigsqcup\nolimits^i X})$. It follows that there is some $c \in B$ with $c \ll \bigsqcup\nolimits^i X$ so that $e \ll' f(c)$. Thus, we obtain that $(i, X) \vdash (c, \{ c \})$, $(c, \{ c \}) \III(f) (e, \{ e \})$ and $(e, \{ e \}) \vdash' S$.
\end{pf}

\begin{lem}\label{lem-funci}
$\fun{\III}{\mathbf{L}}{\mathbf{ISW}}$ is a functor.
\end{lem}

Functoriality follows from Scott continuity. The other property is obvious. 

As we have seen in the preceding section, up to isomorphism every L-domain is generated by an information system with witnesses. Let $\AAA$ and $\AAA'$ be such information systems and $\fun{f}{|\AAA|}{|\AAA'|}$ Scott continuous. If we now construct the information systems corresponding to the domains $|\AAA|$ and $|\AAA'|$ as well as the approximable mapping corresponding to $f$ in the above way and consider the domains and the function generated by these, we will not come back to $f$. The state sets involved will be one level higher up in the power set hierarchy. This can be avoided, however. For $(i, X) \in \Con$ and $a \in A'$, define
\[
(i, X) H^f a \Longleftrightarrow a \in f([X]_i).
\]

\begin{lem}\label{lem-scammod}
$\apmap{H^f}{\AAA}{\AAA'}$.
\end{lem}
\begin{pf}
We have to verify the conditions in Definition~\ref{dn-am}.

(\ref{dn-am-1}) Assume that $(i, X) H^f (k, Y)$ and $(k, Y) \vdash' b$. Then $\{ k \} \cup Y \fsubset f([X]_i)$. Since $f([X]_i)$ is a state, it follows with \ref{st} that $(j, Z) \vdash' (k, Y)$, for some $(j, Z) \in \Con'$ with $\{ j \} \cup Z \subseteq f([X]_i)$. Then $(j, Z) \vdash' b$ and hence $b \in f([X]_i)$, by \ref{dn-st}(\ref{dn-st-2}).

(\ref{dn-am-2}, \ref{dn-am-3}) are obvious by the monotonicity of $f$. (\ref{dn-am-4}) is obvious as well, as is (\ref{dn-am-6}), since $\bt'$ is contained in any state of $A'$, by \ref{dn-st}(\ref{dn-st-2}). 

It remains to verify (\ref{dn-am-5}). Let $(i, X) H^f F$. Then, by Condition~(\ref{st}), there is some $(j, Y) \in \Con'$ with $\{ j \} \cup Y \subseteq f([X]_i)$ and $(j, Y) \vdash' F$. Now, Condition~\ref{dn-infsys}(\ref{dn-infsys-10}) provides us with some $(e, V) \in \Con'$ so that $(j, Y) \vdash' (e, V)$ and $(e, V) \vdash'  F$. It follows with \ref{dn-st}(\ref{dn-st-2}) that $\{ e \} \cup V \subseteq f([X]_i)$. By Lemma~\ref{lem-canba}(\ref{lem-canba-2}) and the Scott continuity of $f$ we therefore obtain that there is some $(k, Z) \in \Con$ with $(i, X) \vdash (k, Z)$ and $\{ e \} \cup V \subseteq f([Z]_k)$. Applying \ref{dn-infsys}(\ref{dn-infsys-10}) again supplies us with some $(c, U) \in \Con$ such that $(i, X) \vdash (c, U)$ and $(c, U) \vdash (k, Z)$. It ensues that $\{ e \} \cup V \subseteq  f([Z]_k) \subseteq f([U]_c)$. Altogether we thus have that $(i, X) \vdash (c, U)$, $(c, U) H^f (e, V)$ and $(e, V) \vdash' F$.
\end{pf}

\begin{lem}\label{lem-surj}
$\LLL(H^f) = f$.
\end{lem}
\begin{pf}
Let $x \in |\AAA|$ and $b \in A'$. Then we have
\begin{align*}
b \in \LLL(H^f)(x) 
&\Longleftrightarrow (\exists (i, X) \in \Con) \{ i \} \cup X \subseteq x \wedge (i, X) H^f b \\
&\Longleftrightarrow (\exists (i, X) \in \Con) \{ i \} \cup X \subseteq x \wedge b \in f([X]_i) \\
&\Longleftrightarrow b \in \bigcup\set{f([X]_i)}{(i, X) \in \Con \wedge \{ i \} \cup X \subseteq x}  \\
&\Longleftrightarrow b \in f(\bigcup\set{[X]_i}{(i, X) \in \Con \wedge \{ i \} \cup X \subseteq x}) \\
&\Longleftrightarrow b \in f(x).
\end{align*}
\end{pf}

Now, conversely, let $\apmap{H}{\AAA}{\AAA'}$. Then $\fun{\LLL(H)}{|\AAA|}{|\AAA'|}$, by Lemma~\ref{lem-apmapst}.

\begin{lem}\label{lem-amscam}
$H^{\LLL(H)} = H$.
\end{lem}
\begin{pf}
For $(i, X) \in \Con$ and $b \in A'$, we have that
\begin{align*}
(i, X) H^{\LLL(H)} b
&\Longleftrightarrow b \in \LLL(H)([X]_i) \notag \\
&\Longleftrightarrow (\exists (j, Y) \in \Con) (i, X) \vdash (j, Y) \wedge (j, Y) H b \\
&\Longleftrightarrow (i, X) H b.
\end{align*}
The last equivalence follows with Lemmas~\ref{lem-amstrong3} and \ref{pn-amint}(\ref{pn-amint-1}), respectively.
\end{pf}

As a consequence of the last two lemmas the functor $\fun{\LLL}{\mathbf{ISW}}{\mathbf{L}}$ is full and faithful. Moreover, by Theorem~\ref{tm-dominsys}, we have that any domain $D$ in $\mathbf{L}$ is isomorphic to $\LLL(A)$, for some information system $A$ in $\mathbf{ISW}$, namely $\III(D)$. With \cite[p.~93, Theorem~1] {macl98} it thus follows that $\LLL$ is an equivalence.

\begin{thm}\label{tm-eqlisw}
The category $\mathbf{ISW}$ of information systems with witnesses and approximable mappings is equivalent to the category $\mathbf{L}$ of L-domains and Scott continuous functions.
\end{thm}

\begin{cor}\label{cor-eqabc}
The categories $\mathbf{aISW}$, $\mathbf{bcISW}$ and $\mathbf{abcISW}$, respectively, of information systems with witnesses satisfying Conditions~(\ref{alg}), (\ref{bc}), or both of them, and approximable mappings are equivalent to the categories $\mathbf{aL}$, $\mathbf{BC}$ and $\mathbf{aBC}$ of algebraic L-domain, bc-domains and algebraic bc-domains with Scott continuous functions.
\end{cor}

\section{Other kinds of information systems}\label{sec-rel}

In this section we will see how some classical types of information systems studied in the literature can be considered as information systems with witnesses in which the witnesses are ignored.

Scott~\cite{sco82} introduced his information systems as a logic-based introduction to countably based algebraic bc-domains. In order to capture the more general continuous bc-domains, Hoofman \cite{ho93} extended Scott's approach and introduced continuous information systems. Moreover, he showed that Scott's information systems or, more exactly, its slight modification introduced by Larsen and Winskel \cite{lw84} are a special case of his information systems.

\subsection{Continuous information systems}\label{ssec-cis}

\begin{dn}\label{dn-cis}
Let $A$ be a set, $\con \subseteq \PPP_f(A)$, and $\vdash \subseteq \con \times A$.  $(A, \con, \vdash)$ is a \emph{continuous information system} if the following conditions hold, for all $a \in A$ and $X, Y \fsubset A$:
\begin{enumerate}
\item\label{dn-cis-1-1}
$\emptyset \in \con$

\item\label{dn-cis-1-2}
$Y \subseteq X \wedge X \in \con \Rightarrow Y \in \con$

\item\label{dn-cis-1-3}
$\{ a \} \in \con$

\item\label{dn-cis-1-4}
$X \vdash Y \Rightarrow Y \in \con$

\item\label{dn-cis-1-5}
$(X, Y \in \con\mbox{} \wedge X \subseteq Y \wedge X \vdash a) \Rightarrow Y \vdash a$

\item\label{dn-cis-1-6}
$(\exists Z \in \con)[X \vdash Z \wedge Z \vdash a] \Leftrightarrow X \vdash a$.
\end{enumerate}
\end{dn}

Set $\Con = A \times \con$ and for $(i, X) \in \Con$ define
\[
(i,X) \vDash a \Longleftrightarrow X \vdash a.
\]

\begin{pn}\label{pn-cisisw}
Let $(A, \con, \vdash)$ be a continuous information system. Then $(A \cup \{ \emptyset \}, \Con,\linebreak \vDash, \emptyset)$ is an information system with witnesses that satisfies Condition~(\ref{bc}).
\end{pn}
\begin{pf}
Condition~(\ref{bc}) is obvious and Conditions~\ref{dn-infsys}(\ref{dn-infsys-1}-\ref{dn-infsys-9}) are easily verified and Condition~\ref{dn-infsys}(\ref{dn-infsys-10}) is a consequence of \cite[Theorem~20]{ho93}.
\end{pf}

As we shall see next, there is a canonical way to pass from an information system with witnesses satisfying Condition~(\ref{bc}) to a continuous information system such that both generate the same domain.

Let $\AAA = (A, \Con, \vdash, \bt)$ be an information system with witnesses so that Requirement~(\ref{bc}) holds. Set $\con = \bigcup\set{\Con(i)}{i \in A}$ and for $a \in A$ and $X \fsubset A$ define
\[
 X \Vdash a \Longleftrightarrow (\exists i \in A) X \in \Con(i) \wedge (i, X) \vdash a.
 \]
 
\begin{lem}\label{lem-bccis}
$\CCC(\AAA) = (A, \con, \Vdash)$ is a continuous information system.
\end{lem} 
\begin{pf}
We have to verify the conditions in Definition~\ref{dn-cis}.

(\ref{dn-cis-1-1})
By Condition~\ref{dn-infsys}(\ref{dn-infsys-1}) we have that $\{ \bt \} \in \Con(\bt)$. Because of \ref{dn-infsys}(\ref{dn-infsys-2}) it follows that $\emptyset \in \Con(\bt)$ and hence that $\emptyset \in \con$.

(\ref{dn-cis-1-2})
Let $Y \in \con$. Then there is some $i \in A$ with $Y \in \Con(i)$. With Condition~\ref{dn-infsys}(\ref{dn-infsys-2}) it follows for $X \subseteq Y$ that $X \in \Con(i)$ as well. Thus, $X \in \con$.

(\ref{dn-cis-1-3})
Since $\{ a \} \in \Con(a)$ by Axiom~\ref{dn-infsys}(\ref{dn-infsys-1}), we have that $\{ a \} \in \con$, for every $a \in A$.

(\ref{dn-cis-1-4})
Let $X \in \con$ and $Y \fsubset A$ with $X \Vdash Y$. Then there is some $i_b \in A$ with $X \in \Con(i_b)$ and $(i_b, X) \vdash b$, for each $b \in Y$. Thus, $X \in \bigcap\set{\Con(i_b)}{b \in Y}$. Because of Condition~(\ref{bc}) we moreover have that for all $c, d \in Y$,
\[(i_c, X) \vdash b \Longleftrightarrow (i_d, X) \vdash b.
\]
If $Y$ is empty, we already know that $Y \in \con$. Otherwise, let $c \in Y$. Then, $(i_c, X) \vdash b$, for all $b \in Y$, that is, $(i_c, X) \vdash Y$. With \ref{dn-infsys}(\ref{dn-infsys-4}) it follows that $Y \in \Con(i_c)$ which shows that $Y \in \con$.

(\ref{dn-cis-1-5})
Let $a \in A$ and $X, Y \in \con$ such that $X \Vdash a$ and $X \subseteq Y$. We must show that $Y \Vdash a$. As a consequence of our assumption there are $i, j \in A$ so that $X \in \Con(i)$, $(i, X) \vdash a$ and $Y \in \Con(j)$. Since $X \subseteq Y$, if follows with Axiom~\ref{dn-infsys}(\ref{dn-infsys-2}) that $X \in \Con(j)$ as well. Thus, $X \in \Con(i) \cap \Con(j)$. Because of (\ref{bc}) we therefore have that $(i, X) \vdash a$, exactly if $(j, X) \vdash a$.
So, $X, Y \in \Con(j)$, $(j, X) \vdash a$ and $X \subseteq Y$ which implies that $(j, Y) \vdash a$, by Axiom~\ref{dn-infsys}(\ref{dn-infsys-5}). Hence, $Y \Vdash a$.
 
(\ref{dn-cis-1-6})
Assume that $X \Vdash Y$ and $Y \Vdash a$. We first need to show that $X \Vdash a$. Since $X \Vdash Y$, there is some $i_b \in A$, for each $b \in Y$, such that $X \in \Con(i_b)$ and $(i_b, X) \vdash b$. Moreover, as $Y \Vdash a$, there is some $j \in A$ with $Y \in \Con(j)$ and $(j, Y) \vdash a$. 
If $Y$ is empty, we have that $\emptyset \Vdash a$ and therefore, by what has just been shown, that $X \Vdash a$. Otherwise, there is some $c \in Y$. Because $X \in \bigcap\set{\Con(i_b)}{b \in Y}$, it follows with (\ref{bc}) that for all $b \in Y$, $(i_b, X) \vdash b$, exactly if $(i_c, X) \vdash b$. Thus, we have that $(i_c, X) \vdash Y$, which by Axiom~\ref{dn-infsys}(\ref{dn-infsys-4}) implies that $Y \in \Con(i_c)$. This shows that $Y \in \Con(j) \cap \Con(i_c)$. With (\ref{bc}) we hence obtain that $(i_c, Y) \vdash a$, exactly if $(j, Y) \vdash a$. Altogether we thus have that $(i_c, X) \vdash Y$ and $(i_c, Y) \vdash a$. Therefore, $(i_c, X) \vdash a$, because of ~\ref{dn-infsys}(\ref{dn-infsys-6}), which means that $X \Vdash a$. 

The converse implication is an easy consequence of Axiom~\ref{dn-infsys}(\ref{dn-infsys-10}).
\end{pf} 
 
Let $\bA = (A, \con, \vdash)$ be a continuous information system. A subset $x$ of $A$ is a \emph{point}, if the following three requirements hold:
\begin{enumerate}
\item\label{pt-1}
$(\forall F \fsubset x) F \in \con$

\item\label{pt-2}
$(\forall X \fsubset x)(\forall a \in A) [X \in \con\mbox{} \wedge X \Vdash a \Rightarrow a \in x]$

\item\label{pt-3}
$(\forall a \in x)(\exists X \fsubset x) X \in \con\mbox{} \wedge X \Vdash a$.

\end{enumerate} 

The collection of points is a continuous bc-domain with respect to set inclusion \cite{ho93} which we also denote by $|\bA|$.

Let $\AAA = (A, \Con, \vdash, \bt)$ be an information system with witnesses such that Condition~(\ref{bc}) holds. Then every state of $\AAA$ is a point of $\CCC(\AAA)$, and conversely. So, both systems generate the same domain.

\begin{pn}\label{pn-infosys-cis}
Let $\AAA$ be an information system with witnesses such that Condition~(\ref{bc}) holds. Then $\CCC(\AAA)$ is a continuous information system such that $|\CCC(\AAA)| = |\AAA|$.
\end{pn}

\subsection{Algebraic information systems}\label{ssec-ais}
 
Next, we consider the algebraic case.

\begin{dn}\label{dn-ais}
Let $A$ be a set, $\bt \in A$, $\con \subseteq \PPP_f(A)$, and $\vdash \subseteq \con \times A$. $(A, \con, \vdash, \bt)$ is an \emph{algebraic information system} if for all $a \in A$ and $X, Y \fsubset A$ the following requirements are satisfied:
\begin{enumerate}
\item\label{dn-ais-1}
$Y \subseteq X \wedge X \in \con \Rightarrow Y \in \con$

\item\label{dn-ais-2}
$\{ a \} \in \con$

\item\label{dn-ais-3}
$X \vdash a \Rightarrow X \cup \{ a \} \in \con$

\item\label{dn-ais-4}
$X \vdash \bt$

\item\label{dn-ais-5}
$(X, Y \in \con\mbox{} \wedge X \vdash Y \wedge X \vdash a) \Rightarrow Y \vdash a$

\item\label{dn-ais-6}
$a \in X \Rightarrow X \vdash a$.

\end{enumerate}
\end{dn}

Every algebraic information system is a continuous information system. Let $\Con$ and $\vDash$ be defined as in Proposition~\ref{pn-cisisw}. 

\begin{pn}\label{pn-aisisw}
Let $(A, \con, \vdash, \bt)$ be an algebraic information system. Then $(A, \Con,\linebreak \vDash, \bt)$ is an information system with witnesses so that Requirements~(\ref{bc}) and (\ref{alg}) hold.
\end{pn}
\begin{pf}
Conditions~\ref{dn-infsys}(\ref{dn-infsys-1}-\ref{dn-infsys-3},\ref{dn-infsys-6}-\ref{dn-infsys-9}) and (\ref{bc}) are obvious, Condition~\ref{dn-infsys}(\ref{dn-infsys-5}) is a consequence of Requirements~\ref{dn-ais}(\ref{dn-ais-5},\ref{dn-ais-6}), and \ref{dn-infsys}(\ref{dn-infsys-4}) follows with \ref{dn-infsys}(\ref{dn-infsys-5}) and \ref{dn-ais}(\ref{dn-ais-3}). With Requirement~\ref{dn-ais}(\ref{dn-ais-6}) we also obtain that Conditions~(\ref{salg}) and hence \ref{dn-infsys}(\ref{dn-infsys-10}) hold.
\end{pf}

For the converse construction we require that the following strengthening (\ref{alg+}) of Condition~(\ref{alg}) holds:
\begin{equation}\tag{ALG+}\label{alg+}
(i, X) \vdash F  \Rightarrow (\exists j \in A) (i, X) \vdash  j \wedge (j, \{ j \}) \vdash j \wedge (j, \{ j \}) \vdash F.
\end{equation}

Note that Condition~(\ref{alg+}) is satisfied by all information systems with witnesses $\III(D)$ generated by algebraic L-domains $D$.

We say that $j \in A$ is \emph{reflexive} if $(j, \{ j \}) \vdash j$, and denote the subset of all such elements in $A$ by $A_{\rm refl}$. 

\begin{lem}\label{lem-refl}
Let $(j, V) \in \Con$ with $\{ j \} \cup V \subseteq A_{\rm refl}$. Then $(j, V)$ is reflexive.
\end{lem}
\begin{pf}
Let $a \in \{ j \} \cup V$. Since $(j, V) \in \Con$, it follows with \ref{dn-infsys}(\ref{dn-infsys-2}) that $\{ a \} \in \Con(j)$. According to our assumption we have that $(a, \{ a \}) \vdash a$. With \ref{dn-infsys}(\ref{dn-infsys-8}) we therefore obtain that also $(j, \{ a \}) \vdash a$. Hence, $(j, V) \vdash a$, because of \ref{dn-infsys}(\ref{dn-infsys-5}). This shows that $(j, V) \vdash (j, V)$.
\end{pf}

Let $(A, \Con, \vdash, \bt)$ be an information system with witnesses such that both Conditions, (\ref{bc}) and (\ref{alg+}), hold. Set 
\begin{gather*}
\con_{\rm refl} = \set{X \fsubset A_{\rm refl}}{(\exists i \in A_{\rm refl}) (i, X) \in \Con} \\
\intertext{and for $a \in A_{\rm refl}$ as well as $X \fsubset A_{\rm refl}$,}
X \vdash_{\rm refl} a \Longleftrightarrow (\exists i \in A_{\rm refl}) (i, X) \in \Con\mbox{} \wedge (i, X) \vdash a.
\end{gather*}
Note that $\bt \in A_{\rm refl}$ because of Axioms~\ref{dn-infsys}(\ref{dn-infsys-3}-\ref{dn-infsys-5}).

\begin{lem}\label{lem-algais}
Let $\AAA = (A, \Con, \vdash, \bt)$ be an information system with witnesses such that Conditions~(\ref{bc}) and (\ref{alg+}) hold. Then
$\RRR(\AAA) = (A_{\rm refl}, \con_{\rm refl}, \vdash_{\rm refl}, \bt)$ is an algebraic information system.
\end{lem}
\begin{pf}
We have to verify the requirements in Definition~\ref{dn-ais}. For \ref{dn-ais}(\ref{dn-ais-1},\ref{dn-ais-2},\ref{dn-ais-5}) the proof proceeds as in Lemma~\ref{lem-bccis}. 

(\ref{dn-ais-3})
Assume that $X \vdash_{\rm refl} a$. Then there is some $i \in A_{\rm refl}$ such that $(i, X) \in \Con$ and $(i, X) \vdash a$. 
As $(i, X)$ is reflexive, by Lemma~\ref{lem-refl}, we have that $(i, X) \vdash \{ i, a \} \cup X$. With Axiom~\ref{dn-infsys}(\ref{dn-infsys-4}) it follows that $X \cup \{ a \} \in \Con(i)$. Therefore, $(i, X \cup \{ a \})$ is reflexive, again by Lemma~\ref{lem-refl}. Thus, $X \cup \{ a \} \in \con$.

(\ref{dn-ais-4})
Since $X \in \con$, there is some $i \in A_{\rm refl}$ so that $(i, X) \in \Con$. By Axiom~\ref{dn-infsys}(\ref{dn-infsys-3}) we have in addition that $(i, \emptyset) \vdash \bt$. With \ref{dn-infsys}(\ref{dn-infsys-5}) therefore obtain that $(i, X) \vdash \bt$, which means that $X \vdash_{\rm refl} \bt$.

(\ref{dn-ais-6})
Suppose that $a \in X$, where $X \in \con_{\rm refl}$. Then $(i, X) \in \Con$, for some $i \in A_{\rm refl}$. With Lemma~\ref{lem-refl} it follows that $(i, X)$ is reflexive. Hence, $(i, X) \vdash a$, that is, $X \vdash_{\rm refl} a$.
\end{pf}

For algebraic information systems the definition of a point can be simplified. Let $\bA = (A, \con, \vdash, \bt)$ be an algebraic information system. A subset $x$ of $A$ is a \emph{point}, if the following two requirements hold:
\begin{enumerate}
\item\label{apt-1}
$(\forall F \fsubset x) F \in \con$

\item\label{apt-2}
$(\forall X \fsubset x)(\forall a \in A) [X \in \con\mbox{} \wedge X \vdash a \Rightarrow a \in x]$.

\end{enumerate}
The collection of points is an algebraic bc-domain with respect to set inclusion \cite{sco82}, and  $\set{\set{a \in A}{X \vdash a}}{X \in \con}$ is its canonical basis. We  denote this domain by $|\bA|$ as well.

Let $\AAA = (A, \Con, \vdash, \bt)$ be an information system with witnesses such that Conditions~(\ref{bc}) and (\ref{alg+}) hold. Then $\RRR(\AAA)$ is an algebraic bc-domain with basis $\set{[X]_i}{(i, X) \in \con_{\rm refl}}$. It follows that the two information systems $\AAA$ and $\RRR(A)$ generate isomorphic domains.

\begin{pn}\label{pn-infosys-ais}
Let $\AAA = (A, \Con, \vdash, \bt)$ be an information system with witnesses such that Conditions~(\ref{bc}) and (\ref{alg+}) hold. Then $\RRR(\AAA)$ is an algebraic information system such that  $|\RRR(\AAA)|$ and $|\AAA|$ are isomorphic domains.
\end{pn}

\section{Other domain logics}\label{sec-domlog}

Similarly to the previous section, we will now investigate the relationship between other logics (not of the information system kind) presented in the literature to characterise L-domains, or subclasses thereof, and  information systems with witnesses.

\subsection{Disjunctive propositional logic}\label{ssec-dpl}

Following Abramsky's programme of developing a Stone-type duality for classes of domains~\cite{ab91}, Chen and Jung~\cite{cj06} introduced \emph{disjunctive propositional logic} and showed that their theories form an algebraic L-domain. Moreover, up to isomorphism each L-domain can be obtained in this way. As we will see, every such logic defines an information system with witnesses in a canonical way.

Formulae of this logic are built out of atomic propositions using binary conjunctions and arbitrary, but provably disjoint, disjunctions. For the proof system sequents as in Gentzen's intuitionistic sequent calculus are employed. These take the form $\Gamma \vdash \phi$ where $\Gamma$ is a finite set of formulae and $\phi$ is a single formula.  As usual the intended meaning is that the conjunction of the propositions in $\Gamma$ entails $\phi$.

\begin{dn}\label{dn-dp}
Let $P$ be a set of elements, the elements of which are called \emph{atomic (disjunctive) propositions)}. Likewise, let $S_{0}$ be a set of sequents of the form $p_{1}, \ldots, p_{n} \vdash F$ where the $p_{i}$ are atomic propositions and $F \in P$ is the syntactic constant for ``false''. Similarly, $T \in P$ is the syntactic constant for ``truth''. The elements of $S_{0}$ are called \emph{atomic disjointness assumptions}, and the pair $(P, S_{0})$ a \emph{disjunctive basis}.  

In addition to the disjointness assumptions, \emph{disjunctive axioms} are allowed which are sequents of the form
\[
p_{1}, \ldots, p_{n} \vdash \dvee_{i \in I} \bigwedge_{j \in M_{i}} q_{j},
\]
where all $p_{k}$ and $q_{j}$ are elements of $P$. Let $S$ be a set which extends $S_{0}$ by such axioms. Note that further to the axioms, $S$ must contain all disjointness assumptions
\[
q_{j_{1}},  \ldots, q_{j_{m}} q_{j'_{1}}, \ldots, q_{j'_{m'}} \vdash F,
\]
for $M_{i} = \{ j_{1}, \ldots, j_{m} \}$ and $M_{i'} = \{ j'_{1}, \ldots, j'_{m'} \}$ with $i \not= i' \in I$.

The class $\LLL(P, S)$ of \emph{disjunctive propositions over $P$ and $S$}, and the class $\bT(P, S)$ of \emph{valid sequents over $P$ and $S$} are generated by mutual transfinite induction according to the following rules:

\begin{itemize}

\item Disjunctive propositions\\

$
\begin{array}{r@{\,\,}l}
\text{(At)} &\dfrac{\phi \in P}{\phi \in \LLL(P, S)} \qquad \qquad
\text{(Const)}\,\, \dfrac{}{T, F \in \LLL(P, S)}  \\[3ex]
\text{(Conj)}
&\dfrac{\phi, \psi \in \LLL(P, S)}{\phi \wedge \psi \in \LLL(P, S)}  \\[3ex]
\text{(Disj)}
&\dfrac{\phi_{i} \in \LLL(P, S)\, (\mathrm{all}\, i \in I) \qquad \phi_{i}, \phi_{j} \vdash F \, (\mathrm{all}\, i \not= j \in P)}{\dvee_{i \in I} \phi_{i} \in \LLL(P, S)}
\end{array}
$

\item Valid sequents\\

$
\begin{array}{r@{\hspace{.3em}}l@{\hspace{3em}}r@{\hspace{.3em}}l}
\text{(Ax)} & \dfrac{(\Gamma \vdash \phi) \in S}{\Gamma \vdash \phi} & \text{(Id)} & \dfrac{\phi \in \LLL(P, S)}{\phi \vdash \phi} \\[3ex]
\text{(Lwk)} & \dfrac{\Gamma \vdash \psi \qquad \phi \in \LLL(P, S)}{\Gamma, \phi \vdash \psi} &
\text{(Cut)} & \dfrac{\Gamma \vdash \phi \qquad \Delta, \phi \vdash \psi}{\Gamma, \Delta \vdash \psi} \\[3ex]
\text{(L$F$)} & \dfrac{\phi \in \LLL(P, S )}{F \vdash \phi} & \text{(R$T$)} & \dfrac{}{\vdash T} \\[4ex]
\text{(L$\wedge$)} & \dfrac{\Gamma, \phi, \psi \vdash \theta}{\Gamma, \phi \wedge \psi}
& \text{(R$\wedge$)} & \dfrac{\Gamma \vdash \phi \qquad \Delta \vdash \psi}{\Gamma, \Delta \vdash \phi \wedge \psi} \\[4ex]
 \text{(L$\dvee$)} & \multicolumn{3}{@{\hspace{.1em}}l}{\dfrac{\Gamma, \phi_{i} \vdash \theta\,\, (\text{all}\, i \in I) \qquad \phi_{i}, \phi_{j} \vdash F\, (\text{all}\, i \not= j \in I)}{\Gamma, \dvee_{i \in I} \phi_{i} \vdash \theta}} \\[4ex]
\text{(R$\dvee$)} & \multicolumn{3}{@{\hspace{.1em}}l}{\dfrac{\Gamma \vdash \phi_{i }\, (\text{some}\, i  \in I) \qquad \phi_{i}, \phi_{j} \vdash F\,\, (\text{all}\, i \not= j \in I)}{\Gamma \vdash \dvee_{i \in I} \phi_{i}}} 
\end{array}
$ 
 
\end{itemize}
\end{dn}

We call a logical system $(P, S)$ of the above kind a \emph{disjunctive sequent calculus}. Note that although the inductive definitions produce proper classes of objects, in each formula the nesting of operators is only finite (though may be unbounded); likewise, the length of any path from assumption to conclusion in a derivation is finite (though a derivation may contain paths of arbitrary length). This is because each rule preserves this property. 
As shown in \cite[Proposition 2.3]{cj06} the logical rules, except (R$\dvee$), can be ``inverted''.   

Given a nonempty finite subset $\Gamma = \{ \phi_{1}, \ldots, \phi_{n} \}$ of $\LLL(P, S )$, we abbreviate the  disjunctive proposition $\phi_{1} \wedge \cdots \wedge \phi_{n}$ as $\bigwedge \Gamma$.

\begin{dn}\label{dn-basdef}
Let $(P, S)$ be a disjunctive logical calculus and $\phi, \psi$ disjunctive propositions over $P$ and $S$.
\begin{enumerate}
\item\label{dn-basdef-1}
$\phi$ is called \emph{contradiction} if $\phi \vdash F$ is derivable. Otherwise, $\phi$ is called \emph{satisfiable}. 

\item\label{dn-basdef-2}
$\phi$ is \emph{completely irreducible} if it is satisfiable and, whenever $\phi \vdash \dvee_{i \in I} \theta_{i}$ can be derived, then there is some $i_{0} \in I$ such that $\phi \vdash \theta_{i_{0}}$ is derivable. 
 
\item\label{dn-basdef-3}
$\phi$ and $\psi$ are \emph{interderivable}, in symbols $\phi \dashv\vdash \phi$,  if both $\phi \vdash \psi$ and $\psi \vdash \phi$ can be derived.

\item\label{dn-basdef-4}
A satisfiable disjunctive proposition built up from atomic propositions and the constant $T$ by conjunctive connectives only is said to be a \emph{simple conjunction}.

\item\label{dn-basdef-5}
A satisfiable disjunctive proposition is called \emph{disjunctive normal form} if it has the form $\dvee_{i \in I} \mu_{i}$, where the $\mu_{i}$ are simple conjunctions that are different from each other.

\item\label{dn-basdef-6}
A disjunctive normal form  $\dvee_{i \in I} \mu_{i}$ is \emph{standard} if each simple conjunction $\mu_{i}$ is completely irreducible.

\end{enumerate}
\end{dn} 

For contradictions $\phi$ we define $F$ to be its standard normal form.
Let $\CCC(P, S )$ and $\ell(P, S )$, respectively, denote the sets of all completely irreducible simple conjunctions and all standard normal forms. (For a proof that $\ell(P, S )$ is actually a set, see~\cite{cj06}.) As is readily verified, interderivability is an equivalence relation.

\begin{lem}\label{lem-uncon}
Let $(P, S)$ be  a disjunctive logical calculus and $\mu \in \CCC(P, S )$. If $\mu \vdash \dvee_{i \in I} \theta_{i}$ is derivable then there is a unique $i_{0} \in I$ such that $\mu \vdash \theta_{i_{0}}$ is derivable. We denote the uniquely determined $\theta_{i_{0}}$ by $\sigma(\mu, \set{\theta_{i}}{i \in I})$.
\end{lem}
\begin{pf}
Let $\mu \vdash \dvee_{i \in I} \theta_{i}$ is derivable. Since $\mu$ is completely irreducible, there is some $i_{0} \in I$ such that $\mu \vdash \theta_{i_{0}}$ can be derived. Suppose that for some $i_{1} \not= i_{0} \in I$ also $\mu \vdash \theta_{i_{1}}$ can be derived. Then there must exist a derivation of $\theta_{i_{0}}, \theta_{i_{1}}  \vdash F$. Consequently, we have a derivation of $\mu \vdash F$, contradicting the satisfiability of $\mu$.  
\end{pf}

\begin{cor}\label{cor-uncon}
Let $(P,  S)$ be  a disjunctive logical calculus, $\mu, \rho \in \CCC(P,  S)$, and $\dvee_{i \in I} \theta_{i}$ a standard normal form so that $\mu \vdash \dvee_{i \in I} \theta_{i}$ is valid. Then the following statements hold:
\begin{enumerate}

\item\label{cor-uncon-1}
$\sigma(\mu, \set{\theta_{i}}{i \in I}) \in \set{\theta_{i}}{i \in I}$. 

\item\label{cor-uncon-2}
$\sigma(\mu, \{ \mu \}) = \mu$

\item\label{cor-uncon-3}
If $\mu \vdash \rho$ and $\rho \vdash \dvee_{i \in I} \theta_{i}$ are valid, then $\sigma(\mu, \set{\theta_{i}}{i \in I}) = \sigma(\rho, \set{\theta_{i}}{i \in I})$.

\end{enumerate}
\end{cor}
\begin{pf}
(\ref{cor-uncon-1}) follows form the definition of $\sigma(\mu, \set{\theta_{i}}{i \in I})$, and (\ref{cor-uncon-2}) holds as $\mu \in \CCC(P,  S)$ and hence $\dvee_{i \in \{ \ast \}}\theta_{i}$ with $\theta_{\ast} = \mu$ is its standard normal form.

(\ref{cor-uncon-3}) Assume that $\rho \vdash \dvee_{i \in I} \theta_{i}$ is valid. Then $\rho \vdash \sigma(\rho, \set{\theta_{i}}{i \in I})$ is derivable. Since $\mu \vdash  \rho$ is valid by assumption, it follows that $\mu \vdash  \dvee_{i \in I} \theta_{i}$ is derivable as well. Thus, $\mu \vdash \sigma(\mu, \set{\theta_{i}}{i \in I})$ is also valid. Note that in addition $\mu \vdash \sigma(\rho, \set{\theta_{i}}{i \in I})$ is valid and $\sigma(\mu, \set{\theta_{i}}{i \in I})$, $\sigma(\rho, \set{\theta_{i}}{i \in I})$ are both contained in $\set{\theta_{i}}{i \in I}$. As a consequence of the preceding lemma, we therefore obtain that $\sigma(\mu, \set{\theta_{i}}{i \in I}) = \sigma(\rho, \set{\theta_{i}}{i \in I})$.
\end{pf}

\begin{dn}\label{dn-Ncalc}
A \emph{disjunctive N-sequent calculus} is a disjunctive sequent calculus in which the constant $T$ is completely  irreducible, and for any satisfiable disjunctive proposition $\phi$ there exists a standard disjunctive normal form  $\dvee_{i \in I} \mu_{i}$ such that
\begin{equation}\label{eq-Ncalc}
\phi \vdash \dvee_{i \in I} \mu_{i} \in \bT(P,  S)\,\,\text{and for all $i \in I$,}\,\, \mu_{i} \vdash \phi \in \bT(P,  S).
\end{equation}
We abbreviate Condition~(\ref{eq-Ncalc}) by $\phi \leftrightharpoons \bigvee_{i \in I} \mu_{i}$ and call the formula $\dvee_{i \in I} \mu_{i}$ \emph{standard normal form of $\phi$}. 
\end{dn}

\begin{pn}\label{pn-unsnf}
Let $\bP$ be a disjunctive N-sequent calculus and $\phi$ a satisfiable disjunctive proposition. If  $\dvee_{i \in I} \mu_{i}$ and $\dvee_{j \in J} \nu_{j}$ of $\phi$ are standard normal forms, then for each $i \in I$ there is a unique $j \in J$ so that $\mu_{i}$ and $\nu_{j}$ are interderivable, and vice versa.
\end{pn}
\begin{pf}
The statement follows as in \cite[Proposition~3.3]{wll20}
\end{pf}

For satisfiable disjunctive propositions $\phi$, when we write $\ip(\phi)$ we mean that $\ip(\phi)$ is a set of completely irreducible simple conjunctions so that $\dvee \ip(\phi)$ is a standard normal form of $\phi$. Here, $\dvee \ip(\phi)$ is an abbreviation for $\dvee_{\rho \in \ip(\phi)} \rho$. Similarly, we write $\ip(\phi) = \Delta$, if $\Delta$ is a set of completely irreducible simple conjunctions so that $\dvee \Delta$ is a standard normal form of $\phi$. In case $\phi$ is a contradiction, we  let $\ip(\phi)$ be the empty set.

\begin{lem}\label{lem-ipinv}
Let $\phi, \phi_{i} \in \LLL(P,  S)$, for $i \in I$ with $\phi_{i}, \phi_{j} \vdash F$, for $i \not= j \in I$. Then the following statements hold:
\begin{enumerate}

\item\label{lem-ipinv-1}
$\ip(\phi)$ is invariant under interderivability.

\item\label{lem-ipinv-2}
$\ip(\dvee_{I \in I} \phi_{i}) = \bigcup_{i \in I} \ip(\phi_{i})$.

\item\label{lem-ipinv-3}
For $\mu \in \CCC(P,  S)$, $\ip(\mu) = \{ \mu \}$.

\end{enumerate}
\end{lem}
\begin{pf}
(\ref{lem-ipinv-1}) is true as interderivable disjunctive propositions have the same standard normal forms.

(\ref{lem-ipinv-2}) For $i \not= j \in I$ let $\mu \in \ip(\phi_{i})$ and $\nu \in \ip(\phi_{j})$. Then the sequents $\mu, \nu \vdash \phi_{i} \wedge \phi_{j} \vdash F$ are derivable. This implies in particular that $\ip(\phi_{i})$ and $\ip(\phi_{j})$ are disjoint. Since in addition we have for distinct $\rho, \kappa \in \ip(\phi_{i})$ that $\rho, \kappa \vdash F$, the formula $\dvee \bigcup_{i \in I} \ip(\phi_{i})$ is a disjunctive proposition over $P$ and $ S$. It remains to show that $\dvee \bigcup_{i \in I} \ip(\phi_{i}) \leftrightharpoons \dvee_{i \in I} \phi_{i}$.

Since $(\ip(\phi_{i}))_{i \in I}$ is a partition of $\bigcup_{i \in I} \ip(\phi_{i})$ it follows with~\cite[Proposition 2.6]{cj06} that 
\[
\dvee \bigcup_{i \in I} \ip(\phi_{i}) \dashv\vdash \dvee_{i \in I} \dvee \ip(\phi_{i}) \dashv\vdash \dvee_{i \in I} \phi_{i}.
\]
Moreover, we have for $\mu \in \bigcup_{i \in I} \ip(\phi_{i})$ that $\mu \in \ip(\phi_{i_{0}})$, for some $i_{0} \in I$, and hence $\mu \vdash \phi_{i_{0}} \vdash \dvee_{i \in I}\phi_{i}$ are derivable sequents. So, $\mu \vdash \dvee_{i \in I}\phi_{i}$ is valid.

(\ref{lem-ipinv-3}) follows as in Corollary~\ref{cor-uncon}(\ref{cor-uncon-2}).
\end{pf}

For what follows let $\bP = (P, S)$ be a disjunctive N-sequent calculus. We will now show how $\bP$ determines an information system $\AAA_{\bP}$ with witnesses so that $\bP$ and $\AAA_{\bP}$ generate isomorphic algebraic L-domains.

Set 
$
A = \CCC(P, S)
$
and for $\mu \in A$ and $\Gamma \fsubset A$ define
\[
(\mu, \Gamma) \in \Con \Longleftrightarrow \mu \vdash \bigwedge \Gamma \text{ is valid}.
\]

According to our choice of $\Gamma$, all $\theta \in \Gamma$ are satisfiable. Hence, $\bigwedge \Gamma$ is satisfiable as well, since otherwise we would have that $\bigwedge \Gamma \vdash F$ is derivable, and thus, because of Rule (L$\wedge$), for all $\theta \in \Gamma$, also $\theta \vdash F$ would be derivable, which is impossible. It follows that $\bigwedge \Gamma \leftrightharpoons \dvee \ip(\bigwedge \Gamma)$. So, $\mu \vdash \bigwedge \Gamma \vdash \dvee \ip(\bigwedge \Gamma)$ and consequently $\mu \vdash \dvee  \ip(\bigwedge \Gamma)$ are derivable.  Therefore, $\sigma(\mu, \ip(\bigwedge \Gamma))$ is defined.
For $\nu \in A$ define
\[
(\mu, \Gamma) \Vdash \nu \Longleftrightarrow \sigma(\mu, \ip(\bigwedge \Gamma)) \vdash \nu \text{ is valid}
\]
and let $\AAA_{\bP} = (A, \Con, \Vdash, T)$.

\begin{pn}\label{pn-seqinfw}
Let $\bP$ be a disjunctive N-sequent calculus. Then $\AAA_{\bP}$ is an information system with witnesses such that 
\begin{enumerate}
\item\label{pn-seqinfw-1}
 All tokens are reflexive.
 
 \item\label{pn-seqinfw-2}
 Condition~\ref{alg+} holds.
 
 \item\label{pn-seqinfw-3}
 If $(\mu, \Gamma), (\mu, \Delta) \in \Con$ then $(\mu, \Gamma \cup \Delta) \in \Con$ as well.
 \end{enumerate}
 \end{pn}
\begin{pf}
Statement~(\ref{pn-seqinfw-1}) is a consequence of Rule~(Id) and Corollary~\ref{cor-uncon}(\ref{cor-uncon-2}). Statement~(\ref{pn-seqinfw-3}) is obvious.
It remains to verify Conditions~\ref{dn-infsys}(\ref{dn-infsys-1}-\ref{dn-infsys-9}) and (\ref{alg+}). Condition~\ref{dn-infsys}(\ref{dn-infsys-10}) is a consequence of (\ref{alg+}).

(\ref{dn-infsys-1}) follows with Rule (Id) for $\bP$.

(\ref{dn-infsys-2}) Assume that $\Delta \subseteq \Gamma$ and $(\mu, \Gamma) \in \Con$. It follows that $\mu \vdash \bigwedge \Gamma$ is derivable. Apply the  inverted Rule (R$\wedge$) to obtain that $\mu \vdash \psi$ is valid, for all $\psi \in \Gamma$, from which it follows with Rule (R$\wedge$) that $\mu \vdash \bigwedge \Delta$ is valid.

 (\ref{dn-infsys-3}) holds because of Rule (R$T$).
 
(\ref{dn-infsys-4}) Suppose that $(\mu, \Gamma) \in \Con$ and $(\mu, \Gamma) \Vdash \Delta$. Then $\mu \vdash \bigwedge \Gamma$ and $\sigma(\mu, \ip(\bigwedge \Gamma)) \vdash \bigwedge \Delta$ are valid. Moreover, by Lemma~\ref{lem-uncon}, the sequents $\mu \vdash \sigma(\mu, \ip(\bigwedge \Gamma)) \vdash \bigwedge \Delta$ are derivable. So, $\mu \vdash \bigwedge \Delta$ is a valid sequent as well, i.e., $(\mu, \Delta) \in \Con$. 

(\ref{dn-infsys-5}) Assume that $(\mu, \Gamma), (\mu, \Delta) \in \Con$ with $\Gamma \subseteq \Delta$ and $(\mu, \Gamma) \Vdash \theta$, where $\theta \in A$. Then $\mu \vdash \dvee \ip(\bigwedge \Delta)$ is derivable and hence $\mu \vdash \sigma(\mu, \ip(\bigwedge \Delta))$ is derivable as well. Observe that $\sigma(\mu, \ip(\bigwedge \Delta)) \in \Delta$. Because $\bigwedge \leftrightharpoons \dvee \ip(\bigwedge)$, we therefore have that $\sigma(\mu, \ip(\bigwedge \Delta)) \vdash \bigwedge \Delta$ and hence also $\sigma(\mu, \ip(\bigwedge \Delta)) \vdash \bigwedge \Gamma$ are derivable; the latter as $\Gamma \subseteq \Delta$. Note that $\sigma(\mu, \ip(\bigwedge \Delta)) \in \CCC(P,  S)$. Hence, the sequents $ \sigma(\mu, \ip(\bigwedge \Delta)) \vdash \sigma(\sigma(\mu, \ip(\bigwedge \Delta)), \ip(\bigwedge \Gamma)) \vdash \dvee \ip(\bigwedge \Gamma)$ are valid. Since also $\mu \vdash \dvee \ip(\bigwedge \Gamma)$ is valid, we have that the sequents $\mu \vdash \sigma(\mu, \ip(\bigwedge \Gamma)) \vdash \dvee \ip(\bigwedge \Gamma)$ are derivable. As a consequence of Corollary~\ref{cor-uncon}(\ref{cor-uncon-3}) it now follows that $\sigma(\mu, \ip(\bigwedge \Gamma)) = \sigma(\sigma(\mu, \ip(\bigwedge \Delta)), \ip(\bigwedge \Gamma))$. Because $\sigma(\mu, \ip(\bigwedge \Gamma)) \vdash \theta$ is valid by assumption, we finally obtain that $\sigma(\mu, \ip(\bigwedge \Delta)) \vdash \theta$ is derivable, i.e., $(\mu, \Delta) \Vdash \theta$.

 (\ref{dn-infsys-6}) Let $(\mu, \Gamma) \in \Con$, $(\mu, \Gamma) \Vdash \Delta$, and $(\mu, \Delta) \Vdash \theta$ with $\theta \in A$. Then $\mu \vdash \bigwedge \Gamma$ and $\mu \vdash \sigma(\mu, \ip(\bigwedge \Gamma)) \vdash \bigwedge\Delta$ are valid sequents. With Corollary~\ref{cor-uncon}(\ref{cor-uncon-3}) we therefore have that $\sigma(\mu, \ip(\bigwedge \Gamma)) =\sigma(\sigma(\mu, \ip(\bigwedge \Gamma)), \ip(\bigwedge \Delta))$. As a further consequence we obtain that $\mu \vdash \bigwedge \Delta$. Then $\sigma(\sigma(\mu, \ip(\bigwedge \Gamma)), \ip(\bigwedge \Delta)) = \sigma(\mu, \ip(\bigwedge \Delta))$, by Lemma~\ref{lem-uncon} and the definition of $ \sigma(\mu, \ip(\bigwedge \Delta))$. With the assumption that $\sigma(\mu, \ip(\bigwedge \Delta)) \vdash \theta$ is derivable it now follows that $\sigma(\mu, \ip(\bigwedge \Gamma)) \vdash \theta$ is derivable, i.e. $(\mu, \Gamma) \Vdash \theta$.

(\ref{dn-infsys-7}) For $\mu, \rho \in A$ suppose that $(\rho, \{ \mu \}) \in \Con$ and let $(\mu, \Gamma) \in Con$. Then $\rho \vdash \mu$ and $\mu \vdash \bigwedge \Gamma$ are valid. Hence, also $\rho \vdash \Gamma$ is valid, which shows that $(\rho, \Gamma) \in \Con$.

(\ref{dn-infsys-8}) Assume that $(\rho, \{ \mu \}) \in \Con$, $(\mu, \Gamma) \in \Con$ and $(\mu, \Gamma) \Vdash \theta$ with $\theta \in A$. Again we have that $\rho \vdash \mu$ and $\mu \vdash \bigwedge \Gamma$ are valid. Moreover, $\sigma(\mu, \ip(\bigwedge \Gamma)) \vdash \theta$ is derivable. Because of Corollary~\ref{cor-uncon}(\ref{cor-uncon-3}), $\sigma(\mu, \ip(\bigwedge \Gamma)) = \sigma(\rho, \ip(\bigwedge))$, which implies that also $\sigma(\rho, \ip(\bigwedge \Gamma)) \vdash \theta$ is derivable, i.e. $(\rho, \Gamma) \Vdash \theta$.

(\ref{dn-infsys-9}) Now, suppose that $(\rho, \{ \mu \}) \in \Con$, $(\mu, \Gamma) \in \Con$ and $(\rho, \Gamma) \Vdash \theta$. Then $\rho \vdash \mu$ and $\mu \vdash \bigwedge \Gamma$ are valid sequents. It follows that also $\rho \vdash \bigwedge \Gamma$ is valid. Since $\sigma(\rho, \ip(\bigwedge \Gamma)) = \sigma(\mu, \ip(\bigwedge \Gamma))$, by Corollar~\ref{cor-uncon}(\ref{cor-uncon-3}), our assumption that $\sigma(\rho, \ip(\bigwedge \Gamma)) \vdash \theta$ is derivable implies that also $\sigma(\mu, \ip(\bigwedge \Gamma)) \vdash \theta$ is derivable. Thus, $(\mu, \Gamma) \Vdash \theta$.

(\ref{alg+}) Let $(\mu, \Gamma) \in \Con$ with $(\mu, \Gamma) \Vdash \Delta$, where $\Delta \fsubset A$. Then $\mu \vdash \Gamma$ is a valid sequent. Set $\nu = \sigma(\mu, \ip(\bigwedge \Gamma))$. Since $\mu \vdash \sigma(\mu, \ip(\bigwedge \Gamma))$ is derivable, we have that $(\mu, \{ \nu \}) \in \Con$. Because of Rule (Id)
$\sigma(\mu, \ip(\bigwedge \Gamma)) \vdash \sigma(\mu, \ip(\bigwedge \Gamma))$ is valid. Thus, $(\mu, \Gamma) \Vdash (\nu, \{ \nu \})$. Since $\mu \vdash \sigma(\mu, \ip(\bigwedge \Gamma))$ is derivable by Lemma~\ref{lem-uncon} and $\sigma(\mu, \ip(\bigwedge \Gamma))$ is completely irreducible, we obtain that $\sigma(\mu, \ip(\nu)) = \sigma(\mu, \ip(\bigwedge \Gamma))$. Thus, $(\mu, \{ \nu \}) \Vdash (\nu, \{ \nu \})$.
By assumption, $\sigma(\mu, \ip(\bigwedge \Gamma)) \vdash \theta$ is derivable, for all $\theta \in \Delta$. Moreover, we have seen that $\mu \vdash \nu$ is valid and therefore, $\sigma(\mu, \{ \nu \}) = \sigma(\nu, \{ \nu \})$, by Corollary~\ref{cor-uncon}(\ref{cor-uncon-3}). Hence, we also have that $(\nu, \{ \nu \}) \Vdash \Delta$.
\end{pf}

For $\phi \in \LLL(P,  S)$ set
\[
[\phi]_{\bP} =  \set{\psi \in \LLL(P, S)}{\psi \dashv\vdash \phi} \cap \ell(P,  S).
\]
The set $\LLL(\bP)$ of all these equivalence classes is called the \emph{Lindenbaum algebra} of $\bP$.

\begin{dn}\label{dn-Dlat}
Let $(L; 0, 1, \sqcap)$ be a meet-semilattice with least element 0 and greatest element 1.
\begin{enumerate}

\item\label{dn-Dlat-1} For $x, y \in L$ we say that $x$ and $y$ are \emph{disjoint} if $x \cap y = 0$.

\item\label{dn-Dlat-2} A subset $B$ of $L$ is \emph{disjoint} if each pair of distinct elements $x$ and $y$ in $B$ are disjoint.

\item\label{dn-Dlat-3} The semilattice is called \emph{disjunctive semilattice} if every disjoint subset has a supremum. Joins of disjoint subsets $B$ are denoted by $\sqdvee B$.

\item\label{dn-Dlat-4} A disjunctive semilattice $L$ is said to be \emph{distributive} if
\[
x \sqcap (\sqdvee B) = \sqdvee_{y \in B} x \sqcap y
\]
holds for each element $x \in L$ and disjoint subset $B$ of $L$.

\item\label{dn-Dlat-5} An element $u$ of a disjunctive semilattice $L$ is a \emph{disjunctively completely coprime} if $u \sqsubseteq \sqdvee B$ always implies $u \sqsubseteq v$ for some element $v$ of the disjoint subset $B$ of $L$. 

\item\label{dn-Dlat-6} The disjunctive semilattice $L$ is called \emph{stable} if every element $x$ is the disjoint supremum of disjunctively completely coprimes and the top element $1$ is a disjunctively completely coprime.

\end{enumerate}
\end{dn}

\begin{dn}\label{dn-cpfil}
A \emph{disjunctive completely prime filter} of $\LLL(\bP)$ is a subset that is closed under finite meets and inaccessible by disjoint suprema.
\end{dn}

Note that the empty set is deemed to be disjoint in the present framework, so a disjunctive completely prime filter cannot contain the least element $0$ of a disjunctive semilattice. Conversely, the greatest element $1$ is always a member.

\begin{thm}[Chen, Jung~\cite{cj06}]\label{thm-cj}
\begin{enumerate}

\item\label{thm-cj-1}
The set of all disjunctive completely prime filter of a stable semilattice forms an algebraic L-domain $\FF$ when ordered by set inclusion, and vice versa. 

\item\label{thm-cj-2}
Let $\bP$ be a disjunctive N-sequent calculus. Then the Lindenbaum  algebra $\LLL(\bP)$ of $\bP$ is a stable semilattice, where the lattice operations are obtained via the corresponding logical connectives applied to representatives, and thus the equivalence classes of completely irreducible simple conjunctions are exactly the disjunctively completely coprimes of $\LLL(\bP)$.

\end{enumerate}
\end{thm}

We will next investigate the connection between states of $\AAA_{\bP}$ and disjunctive completely prime filters of $\LLL(\bP)$. For states $x$ of $\AAA_{\bP}$ set
\[
\FFF(x) = \set{[\phi]_{\bP} \in \LLL(\bP)}{\ip(\phi) \cap x \not= \emptyset}
\]

\begin{lem}\label{lem-stfil}
$\FFF(x)$ is a disjunctive completely prime filter of $\LLL(\bP)$, for every  $x \in |\AAA_{\bP}|$. 
\end{lem}
\begin{pf}
Let $x \in |\AAA_{\bP}|$. We have to show that $\FFF(x)$ is closed under (1) finite meets and (2) inaccessible by disjoint suprema.

(1) Let $[\phi]_{\bP}, [\psi]_{\bP} \in \FFF(x)$. Then there exist $\rho \in \ip(\phi) \cap x$ and $\nu \in \ip(\psi) \cap x$. By Proposition~\ref{pn-stsing} there are thus $\Gamma \fsubset x$ and $\mu \in x$ so that $(\mu, \Gamma) \in \Con$ and $(\mu, \Gamma) \Vdash \{ \rho, \nu \}$. It follows that 
\[
\mu \vdash \sigma(\mu, \ip(\bigwedge \Gamma)) \vdash \rho \wedge \nu \vdash \phi \wedge \psi
\]
are valid sequents. Hence, $\mu \vdash \phi \wedge \psi$ is valid as well, from which we obtain that
\[
\mu \vdash \sigma(\mu, \ip(\phi \wedge \psi)) \vdash \phi \wedge \psi
\]
and therefore
\[
\sigma(\mu, \ip(\bigwedge \Gamma)) \vdash \sigma(\sigma(\mu, \ip(\bigwedge \Gamma)), \ip(\phi \wedge \psi)) \vdash \psi \wedge \psi
\]
are derivable. By Corollary~\ref{cor-uncon}(\ref{cor-uncon-3}), $\sigma(\sigma(\mu, \ip(\bigwedge \Gamma)), \ip(\phi \wedge \psi)) = \sigma(\mu, \ip(\phi \wedge \psi))$. Thus, we have that $(\mu \Gamma) \Vdash   \sigma(\mu, \ip(\phi \wedge \psi))$, which by Condition~\ref{dn-st}(\ref{dn-st-2}) implies that $\sigma(\mu, \ip(\phi \wedge \psi)) \in x$. Since, by Corollary~\ref{cor-uncon}(\ref{cor-uncon-1}), $\sigma(\mu, \ip(\phi \wedge \psi)) \in \ip(\phi \wedge \psi)$, we have that $\ip(\phi \wedge \psi) \cap x$ is not empty, i.e., $[\phi]_{\bP} \cap [\psi]_{\bP} \cap x = [\phi \wedge \psi]_{\bP} \cap x \not= \emptyset$.

(2) Let $(\phi_{i})_{i \in I} \subseteq \LLL(P,  )$ with $\phi_{i}, \phi_{j} \vdash F$, for $i \not= j \in I$, and assume that $\bigsqcup_{i \in I} [\phi_{i}]_{\bP} \in \FFF(x)$. Since  $\bigsqcup_{i \in I} [\phi_{i}]_{\bP} = [\dvee_{i \in I} \phi_{i}]_{\bP}$ it follows that $\ip(\dvee_{i \in I} \phi_{i})$ intersects $x$. As we have seen in Lemma~\ref{lem-ipinv}(\ref{lem-ipinv-2}), $\ip(\dvee_{i \in I} \phi_{i}) = \bigcup_{i \in I} \ip(\phi_{i})$. Hence, for some $i_{0} \in I$, $\ip(\phi_{i_{0}})$ intersects $x$, which means that $[\phi_{i_{0}}]_{\bP} \in x$.
\end{pf}

Now, conversely, let $\PPP$ be a disjunctive completely prime filter of $\LLL(\bP)$ and set
\[
\SSS(\PPP) = \set{\mu \in \CCC(P,  )}{[\mu]_{\bP} \in \PPP}.
\]

\begin{lem}\label{lem-filst}
$\SSS(\PPP) \in |\AAA_{\bP}|$, for all disjunctive completely prime filters of $\LLL(\bP)$.
\end{lem}
\begin{pf}
As shown in \cite{cj06}, every disjunctive completely prime filter of $\LLL(\bP)$ is the directed union of all up-sets $\up c$, with $c$ a disjunctively completely coprime in $\PPP$. Therefore it suffices to demonstrate that for all $\mu \in \CCC(P,  )$, $\SSS(\up [\mu]_{\bP})$ is a state of $\AAA(\bP)$. 

Let $\rho \in \CCC(P,  )$. Then we obtain with Corollary~\ref{cor-uncon}(\ref{cor-uncon-2}) that
\[
\rho \in \SSS(\up [\mu]_{\bP}) \Leftrightarrow [\rho]_{\bP} \sqsupseteq [\mu]_{\bP} \Leftrightarrow \mu \vdash \rho \Leftrightarrow \sigma(\mu, \{ \mu \}) \vdash \rho \Leftrightarrow (\mu, \{ \mu \}) \Vdash \rho \Leftrightarrow \rho \in [\{ \mu \}]_{\mu},
\]
which shows that $\SSS(\up [\mu]_{\bP}) = [\{ \mu \}]_{\mu} \in |\AAA(\bP)|$.
\end{pf}

So far we have shown that $\fun{\FFF}{|\AAA_{\bP}|}{\FF_{\bP}}$ and $\fun{\SSS}{\FF_{\bP}}{|\AAA_{\bP}|}$, where $\FF_{\bP}$ is the algebraic L-domain formed by the disjunctively completely prime filters of $\LLL(\bP)$.  In the remainder of this section we will demonstrate that both maps are Scott continuous and invers to each other. 

\begin{lem}\label{lem-fcont}
 $\fun{\FFF}{|\AAA_{\bP}|}{\FF_{\bP}}$ is Scott continuous.
 \end{lem}
 \begin{pf}
 Obviously $\FFF$ is monotone. Therefore it suffices to show for directed set $E \subseteq |\AAA_{\bP}|$  that  $\FFF(\bigcup E) \subseteq \bigcup \set{\FFF(x)}{x \in E}$.
 
 Let $[\phi]_{\bP} \in \FFF(\bigcup E)$. Then $\ip(\phi)$ intersects $\bigcup E$. Thus, there is some $x \in E$ so that $\ip(\phi)$ intersects $x$, which shows that $\phi \in \FFF(x) \subseteq \bigcup\set{\FFF(y)}{y \in E}$.
 \end{pf}
 
 \begin{lem}\label{lem-scont}
 $\fun{\SSS}{\FF_{\bP}}{|\AAA_{\bP}|}$ is Scott continuous.
 \end{lem}
 \begin{pf}
Monotonicity is obvious. It remains to show that for directed $\EE \subseteq \FF_{\bP}$, $\SSS(\bigcup \EE) \subseteq \bigcup\set{\SSS(\QQQ)}{\QQQ \in \EE}$.

Let $\mu \in \SSS(\bigcup \EE)$. Then $[\mu]_{\bP} \in \bigcup \EE$. Therefore, there is some $\PPP \in \EE$ with $[\mu]_{\bP}$, which shows that $\mu \in \SSS(\PPP) \subseteq \bigcup\set{\SSS(\QQQ)}{\QQQ \in \EE}$.
 \end{pf}

 \begin{lem}\label{lem-fsid}
 \begin{enumerate}
 
 \item\label{lem-fsid-1}
 $\SSS \circ \FFF = \id_{|\AAA_{\bP}|}$
 
 \item\label{lem-fsid-2}
$\FFF \circ \SSS = \id_{\FF_{\bP}}$.

\end{enumerate}
\end{lem}
\begin{pf}
(\ref{lem-fsid-1})  Let $x \in |\AAA_{\bP}|$ and $\mu \in \CCC(P,  )$. Then we obtain with Lemma~\ref{lem-ipinv}(\ref{lem-ipinv-3}) that
\[
\mu \in \SSS(\FFF(x)) \Longleftrightarrow [\mu]_{\bP} \in \FFF(x)
\Longleftrightarrow \ip(\mu) \cap \not= \emptyset
\Longleftrightarrow \mu \in x.
\]

(\ref{lem-fsid-2}) Now, let $\phi \in \LLL(P,  )$ and $\PPP \in \FF_{\bP}$. Then we have
\begin{align*}
\phi_{\bP} \in \FFF(\SSS(\PPP)) \Longleftrightarrow \ip(\phi) \cap \SSS(\PPP) \not= \emptyset
&\Longleftrightarrow (\exists \rho \in \ip(\phi))\, \rho \in \SSS(\PPP) \\
&\Longleftrightarrow (\exists \rho \in \ip(\phi))\, [\rho]_{\bP} \in \PPP
\Longleftrightarrow [\phi]_{\bP} \in \PPP.
\end{align*}
For the proof of the last equivalence assume that $\rho \in \ip(\phi)$ with $[\rho]_{\bP} \in \PPP$. Then $\rho \vdash \phi$ is derivable, by the definition of $\ip(\phi)$. Thus, $[\rho]_{\bP} \sqsubseteq [\phi]_{\bP}$. Since disjunctive completely prime filters are closed under the lattice order, it follows that also $[\phi]_{\bP} \in \PPP$.

Conversely, let $[\phi]_{\bP} \in \PPP$. Since $\phi \leftrightharpoons \dvee \ip(\phi)$, we have that $[\phi]_{\bP} = \sqdvee_{\nu \in \ip(\phi)_{\bP}} [\nu]_{\bP}$. As a disjunctive completely prime filter $\PPP$ is inaccesible by disjoint suprema. Therefore, there is some $\rho \in \ip(\phi)$ with $[\rho]_{\bP} \in \PPP$.
\end{pf}

Let us now summarise what we have obtained in this section.
\begin{thm}\label{thm-maincj}
Let $\bP$ be a disjunctive N-sequent calculus. Then an information system $\AAA_{\bP}$ with witnesses can be constructed so that 
\begin{enumerate}
\item \label{thm-maincj-1}
All elements of $A$ are reflexive.

\item \label{thm-maincj-2}
Condition~\ref{alg+} holds. 

\item \label{thm-maincj-3}
If $(a, X), (a, Y) \in \Con$, then $(a, X \cup Y) \in \Con$ as well.

\item \label{thm-maincj-4}
The L-domain $|\AAA_{\bP}|$ associated with $\AAA_{\bP}$ is isomorphic to the L-domain  $\FF_{\bP}$ generated by the Lindenbaum algebra $\LLL(\bP)$ of $\bP$.
\end{enumerate}
\end{thm}

For the converse construction of an N-sequent calculus $\bP_{\AAA}$ from a given  information system with witnesses such that both generate isomorphic L-domains, let $\AAA = (A, \Con, \vdash, \bt)$ be an information system with witnesses so that all tokens in $A$ are reflexive, Condition~(\ref{alg+}) holds, and with $(a, X), (a, Y)
\in \Con$ also $(a, X \cup Y) \in \Con$.

Then $\bP_{\AAA}$ is defined as follows:
\begin{itemize}
 \item Atomic disjunctive propositions
 \[
 F,\, p_{(a, X)},
 \]
for all $(a, X) \in \Con$. We use $p_{(\bt, \{ \bt \})}$ as syntactic constant for ``truth''.

\item Atomic disjointness assumptions
\[
p_{(a_{1}, X_{1})}, \ldots, p_{(a_{n}, X_{m})} \vDash F,
\]
for all $(a_{1}, X_{n}) \ldots, (a_{n}, X_{n}) \in \Con$ $(n > 0)$ such that for no $c \in A$, $(c, \{ a_{1}, \ldots, a_{n} \}) \in \Con$. 

 \item  Disjunctive axioms
 
\begin{itemize}
\item 
$
p_{(a, X)} \vDash p_{(b, \{ b \})},
$
\quad where $(a, X) \in \Con$ and $(a, X) \vdash b$.

\item
$p_{(a, X_{1})}, \ldots, p_{(a, X_{n})} \vDash p_{(a, \bigcup_{i=1}^{n} X_{i})}$, \qquad
for all $(a, X_{1}), \ldots, (a, X_{n}) \in \Con$.
\end{itemize}
\end{itemize}

Let $(a_{1}, X_{1}), \ldots, (a_{n}, X_{n}) \in \Con$ and set 
\[
A(a_{1}, \ldots, a_{n}) = \set{a \in  A}{(a, \{ a_{1}, \ldots, a_{n} \}) \in \Con}.
\]
Then we have for $a, b \in  A(a_{1}, \ldots, a_{n}))$ that $(a, X_{i}), (b, X_{i}) \in \Con$, for $1 \le i \le n$, and hence that $(a, \bigcup_{i=1}^{n}X_{i}), (b, \bigcup_{i=1}^{n}X_{i}) \in \Con$. Define
$
a \xsim b, 
$
if for some $c \in A(a_{1}, \ldots, a_{n})$, $(a, \{ c \}), (b, \{ c \}) \in \Con$ or $(c, \{ a, b \}) \in \Con$, and let $\hxsim$ be the corresponding transitive closure. Then $\hxsim$ is an equivalence relation. Let $R((a_{1}, X_{1}), \ldots, (a_{n}, X_{n}))$ be a system of representatives, and assume for $d_{1}, d_{2} \in R((a_{1}, X_{1}), \ldots, (a_{n}, X_{n}))$ that there is some $e \in A(a_{1}, \ldots, a_{n})$ such that $(e, \{ d_{1}, d_{2} \}) \in \Con$ or $(d_{1}, \{ e \}), (d_{2}, \{ e \}) \in \Con$. It follows that $d_{1} \xsim d_{2}$ and hence $d_{1} \hxsim d_{2}$, which is impossible by the choice of $d_{1}$ and $d_{2}$.

\begin{lem}\label{lem-rep}
Let $(a_{1}, X_{1}), \ldots, (a_{n}, X_{n}) \in \Con$. Then the following statements hold:
\begin{enumerate}
\item\label{lem-rep-1}
For $d_{1}, d_{2} \in R((a_{1}, X_{1}), \ldots, (a_{n}, X_{n}))$ with , $p_{(d_{1}, \bigcup_{i=1}^{n} X_{i})}, p_{(d_{2}, \bigcup_{i=1}^{n} X_{i})} \vDash F$ is valid.

\item\label{lem-rep-2}
For $a, b \in A(a_{1}, \ldots, a_{n})$ with $a \hxsim b$, $p_{(a, \bigcup_{i=1}^{n} X_{i})}$ and $p_{(b, \bigcup_{i=1}^{n} X_{i})}$ are interderivable.

\end{enumerate}
\end{lem}
\begin{pf}
It remains to derive Statement~(\ref{lem-rep-2}). If $a \hxsim b$, then $[\bigcup_{i=1}^{n} X_{i}]_{a} = [\bigcup_{i=1}^{n} X_{i}]_{b}$, because of Conditions~\ref{dn-infsys}(\ref{dn-infsys-7},\ref{dn-infsys-8}). Since all elements in $A$ are reflexive, it follows with Lemma~\ref{lem-refl} that also $(a, \bigcup_{i=1}^{n} X_{i})$ is reflexive. Thus, $\{ a \} \cup \bigcup_{i=1}^{n} X_{i} \subseteq [\bigcup_{i=1}^{n} X_{i}]_{a} = [\bigcup_{i=1}^{n} X_{i}]_{b}$, which implies that $(b, \bigcup_{i=1}^{n} X_{i}) \vdash (a, \bigcup_{i=1}^{n} X_{i})$. By the disjunctive axioms we therefore obtain that $p_{(b, \bigcup_{i=1}^{n} X_{i})} \vDash p_{(a, \{ c \})}$ is valid, for all $c \in \bigcup_{i=1}^{n} X_{i}$. Consequently, also $p_{(b, \bigcup_{i=1}^{n} X_{i})} \vDash p_{(a, \bigcup_{i=1}^{n} X_{i})}$ is valid. Similarly, it follows that also $p_{(a, \bigcup_{i=1}^{n} X_{i})} \vDash p_{(b, \bigcup_{i=1}^{n} X_{i})}$ is derivable.
\end{pf}

\begin{lem}\label{lem-witNseq} 
\begin{enumerate}
\item\label{lem-witNseq-1}
$\bigwedge_{i=1}^{n} p_{(a_{i}, X_{i})} \Dashv \vDash p_{(a, \bigcup_{{i =1}}^{n} X_{i})}$, for all $a \in A$ so that $(a, \{ a_{1}, \ldots, a_{n} \}) \in \Con$. 

\item\label{lem-witNseq-2}
For all $(a, X) \in \Con$, $p_{(a, X)}$ is completely irreducible.

\item\label{lem-witNseq-3}
For $(a_{1}, X_{1}), \ldots, (a_{n}, X_{n}) \in \Con$, 
\[
\ip(\bigwedge_{i=1}^{n} p_{(a_{i}, X_{i})}) = \set{p_{(c, \bigcup_{i=1}^{n} X_{i})}}{c \in R((a_{1}, X_{1}), \ldots, (a_{n}, X_{n}))}.
\]

\item\label{lem-witNseq-4}
$\bP$ is a disjunctive N-sequent calculus.

\end{enumerate}
\end{lem} 
\begin{pf}
(\ref{lem-witNseq-1}) is an immediate consequence of the disjunctive axioms and  Rules~(L$\wedge$) as well as its invert.

(\ref{lem-witNseq-2}) Assume that $p_{(a, X)} \vDash \dvee_{i \in I} \phi_{i}$ is valid. Because of the form of the disjunctive axioms, Rule~(R$\dvee$) must have been applied in the derivation. So, there is some $i_{0} \in I$ such that $p_{(a, X)} \vDash \phi_{i_{0}}$ is derivable, as was to be shown.

(\ref{lem-witNseq-3}) is a consequence of the first statement and Lemma~\ref{lem-rep}(\ref{lem-rep-1}).

(\ref{lem-witNseq-4})
Let $\phi \in \LLL(P, S)$. If $\phi$ is a contradiction, $\phi \vDash F$ is a valid sequent. Moreover, $F \vDash \phi$ is valid by Rule~(L$F$). In case $\phi$ is satisfiable, it follows with \cite[Theorem~2.8]{cj06} that $\phi$ is interderivable with a formula of the form $\dvee_{i \in I} \bigwedge_{j \in M_{i}} p_{(a_{j}^{i}, X_{j}^{i})}$, where the $M_{i}$ are finite and different from each other. As in Lemma~\ref{lem-ipinv}(\ref{lem-ipinv-2}) we finally obtain that $\dvee_{i \in I} \bigwedge_{j \in M_{i}}  p_{(a^{i}_{j}, X^{i}_{j})}$ is interderivable with $\dvee \bigcup_{i \in I} \ip(\bigwedge_{j \in M_{i}}  p_{(a^{i}_{j}, X^{i}_{j})})$.
 
Moreover, if $p_{(b, Y)} \in \bigcup_{i \in I} \ip(\bigwedge_{j \in M_{i}}  p_{(a^{i}_{j}, X^{i}_{j})})$, then there is a valid sequent 
\[
p_{(b, Y)} \vDash \bigwedge_{j \in M_{i_{0}}}  p_{(a^{i_{0}}_{j}, X^{i_{0}}_{j})},
\]
for some $i_{0} \in I$. Hence, also $p_{(b, Y)} \vDash \dvee_{i \in I} \bigwedge_{j \in M_{i}}  p_{(a^{i}_{j}, X^{i}_{j})}$ is derivable by Rule~(R$\dvee$), and thus  $p_{(b, Y)} \vDash \phi$, because of the interderivbility of the latter right hand sides.
\end{pf}

As shown in Theorem~\ref{thm-maincj}, every disjunctive N-sequent calculus $\bQ$ gives rise to an information system with witnesses $\BBB_{\bQ}$ such that both generate isomorphic L-domains. Let $\BBB_{\bP_{\AAA}} = (P_{\AAA}, \Con_{\bP_{\AAA}}, \Vvdash, p_{(\bt, \{ \bt \})})$ with $P_{\AAA}$ the set of atomic propositions of $\bP_{\AAA}$,  the information system with witnesses corresponding to the calculus $\bP_{\AAA}$ just defined. Then we have for $(b, Y), (a_{1}, X_{1}), \ldots, (a_{n}, X_{n}) \in \Con$ that
\begin{align*}
&(p_{(b, Y)}, \{ p_{(a_{1}, X_{1})}, \ldots, p_{(a_{n}, X_{n})} \}) \in \Con_{\bP_{\AAA}} \\
&\qquad\Longleftrightarrow p_{(b, Y)} \vDash \bigwedge_{i=1}^{n} p_{(a_{i}, X_{i})} \text{ is valid} \\
&\qquad\Longleftrightarrow p_{(b, Y)} \vDash p_{(a_{i}, X_{i})} \text{ is valid, for all $1 \le i \le n$} \\
&\qquad\Longleftrightarrow (b, Y) \vdash (a_{i}, X_{i})\text{, for all $1 \le i \le n$}.
\end{align*}
Hence, $(b, \{ a_{1}, \ldots, a_{n} \} ), (b, \bigcup_{i=1}^{n} X_{i}) \in \Con$, which implies that 
\[
\sigma(p_{(b, Y)}, \ip(\bigwedge_{i=1}^{n} p_{(a_{i}, X_{i})})) = p_{(c, \bigcup_{i=1}^{n} X_{i})},
\]
for some $c \in R((a_{1}, X_{1}), \ldots, (a_{n}, X_{n}))$ with $c \hxsim b$.
Moreover, we have for $(d, Z) \in \Con$ that
\begin{align*}
&(p_{(b, Y)}, \{\, p_{(a_{i}, X_{i})} \mid 1 \le i \le n \,\}) \Vvdash p_{(d, Z)} \\
&\qquad\Longleftrightarrow p_{(c, \bigcup_{i=1}^{n} X_{i})} \vDash p_{(d, Z)} \text{ is valid} \\
&\qquad\Longleftrightarrow p_{(b, \bigcup_{i=1}^{n} X_{i})} \vDash p_{(d, Z)} \text{ is valid} \qquad \text{(by Lemma~\ref{lem-rep}(\ref{lem-rep-2}))} \\
&\qquad\Longleftrightarrow (b, \bigcup_{i=1}^{n} X_{i}) \vdash (c, Z).
\end{align*}

\begin{lem}\label{lem-sp}
For $\bx \in |\BBB_{\bP_{\AAA}}|$ let 
$
\dx(\bx) = \bigcup \set{\{ c \} \cup Z}{p_{(c, Z)} \in \bx}.
$
Then $\dx(\bx) \in |\AAA|$.
\end{lem}
\begin{pf}
It suffices to verify Conditions~(\ref{st}) and \ref{dn-st}(\ref{dn-st-2}). 

For Condition~(\ref{st}) let $G \fsubset x$ and define $\GG = \set{p_{(c, Z)}}{\{ c \} \cup Z \fsubset x}$. Then $\GG \fsubset \bx$. Hence, there are $p_{(b, Y)}, p_{(a_{1}, X_{1})}, \ldots, p_{(a_{n}, X_{n})} \in \bx$ so that $(p_{(b, Y)}, \set{p_{(a_{i}, X_{i})}}{1 \le i \le n}) \in \Con_{\BBB_{\bP_{\AAA}}}$ and $(p_{(b, Y)}, \set{p_{(a_{i}, X_{i})}}{1 \le i \le n}) \Vvdash \GG$. It follows that $\{ b, a_{1}, \ldots, a_{n} \} \cup Y \cup \bigcup_{i=1}^{n} X_{i} \subseteq x$ and $(b, \bigcup_{i=1}^{n} X_{i}) \vdash G$.

For Condition~\ref{dn-st}(\ref{dn-st-2}) let $(b, X) \in \Con$ so that $\{ b \} \cup X \fsubset x$ and $(b, X) \vdash a$. Then $p_{(b, X)} \in \bx$ and $(p_{(b, X)}, \{ p_{(b, X)} \}) \Vvdash p_{(a, \{ a \})}$. Hence, $p_{(a, \{ a \})} \in \bx$ and therefore $a \in x$.
\end{pf}

The next result is obtained similarly.
\begin{lem}\label{lem-ps}
For $x \in |\AAA|$ let 
\[
\pd(x) = \bigcup\set{\{ p_{(c, Z)} \} \cup \set{p_{(a_{i}, X_{i})}}{1 \le i \le n}}{ \{ c, a_{1}, \ldots, a_{n} \} \cup Y \cup \bigcup_{i=1}^{n} X_{i} \fsubset \bx}.
\]
Then $\pd(x) \in |\BBB_{\bP_{\AAA}}|$.
\end{lem}

As is readily verified the maps $\fun{\dx}{|\BBB_{\bP_{\AAA}}|}{|\AAA|}$ and $\fun{\pd}{|\AAA|}{|\BBB_{\bP_{\AAA}}|}$ are Scott continuous and invers to each other. It follows that the L-domains $|\AAA|$ and $|\BBB_{\bP_{\AAA}}|$ are isomorphic. As we have seen in Theorem~\ref{thm-maincj}, the L-domains $\FF_{\bP_{\AAA}}$ generated by the Lindenbaum algebra of $\bP_{\AAA}$ and $|\BBB_{\bP_{\AAA}}|$ are isomorphic as well. So, $|\AAA|$ and $\FF_{{\bP_\AAA}}$ are isomorphic.

\begin{thm}\label{thm-inftoprop}
Let $\AAA = (A, \Con, \vdash, \bt)$ be an information system with witnesses so that 
\begin{enumerate}

\item \label{thm-inftoprop-1}
All tokens are reflexive.

\item \label{thm-inftoprop-2}
Condition~\ref{alg+} holds. 

\item \label{thm-inftoprop-3}
If $(\mu, \Gamma), (\mu, \Delta) \in \Con$, so is $(\mu, \Gamma \cup \Delta)$.

\end{enumerate}
Then a disjunctive N-sequent calculus $\bP_{\AAA}$ can be constructed such that the L-domain $\FF_{\bP_{\AAA}}$ generated by the Lindenbaum algebra $\LLL(\bP_{\AAA})$ of $\bP_{\AAA}$ is isomorphic to the L-domain $|\AAA|$  associated with $\AAA$.
\end{thm}

\subsection{Finitary disjunctive propositional logic}

Wang and Li~\cite{wll20} used a finitary fragment of disjunctive propositional logic to obtain results of the kind as in Chen and Jung~\cite{cj06} for Lawson compact algebraic L-domains. The logic is finitary in as much as it only contains a binary connective for disjoint disjunctions.

\begin{dn}\label{dn-fdp}
Let $(Q, R_{0})$ be a disjunctive basis. The class $\LLL_{f}(Q, R_{0})$ of \emph{finitary disjunctive propositions over $Q$ and $R_{0}$} and the class $\bT_{f}(Q, R_{0})$ of \emph{finitary valid sequents over $Q$ and $R_{0}$} are generated by mutual induction by using
\begin{itemize}

\item Finitary disjunctive propositions\\

Rules (At), (Const), (Conj) as in Definition~\ref{dn-dp}, and \\

$
\begin{array}{r@{\,\,}l}
\text{(Disj)} &  \dfrac{\phi, \psi \in \LLL_{f}(P,  ) \qquad \phi, \psi \vdash F}{\phi \svee \psi \in \LLL_{f}(P,  )}
\end{array}
$

\item Finitary valid sequents\\

Rules (Ax), (Id), (Lwk), (Cut), (L$F$), (R$T$), (L$\wedge$), (R$\wedge$) as in Definition~\ref{dn-dp}, and\\

$
\begin{array}{r@{\hspace{.3em}}l@{\hspace{3em}}r@{\hspace{.3em}}l}
\text{(L$\svee$)} & \dfrac{\Gamma, \phi \vdash \theta \qquad \Gamma, \psi \vdash \theta \qquad \phi, \psi \vdash F}{\Gamma, \phi \svee \psi \vdash \theta}
& \text{(R$\svee$)} & \dfrac{\Gamma \vdash \phi \qquad \phi, \psi \vdash F}{\Gamma \vdash \phi \svee \psi}
\end{array}
$
\end{itemize}
\end{dn}

Note that definitions and results of the preceding section carry over to the present case with only minor changes.  Instead of complete irreducibility one has to assume  irreducibility.

\begin{dn}\label{dn-irr}
A satisfiable finitary disjunctive proposition $\mu$ is said to be \emph{irreducible} if, whenever $\mu \vdash \phi \svee \psi$ is valid, then $\mu \vdash \phi$ or $\mu \vdash \psi$ is valid. We denote the set of all irreducible simple conjunctions by $\CCC_{f}(Q, R_{0})$.  
\end{dn}

If the logical system just introduced satisfies the requirements of Definition~\ref{dn-Ncalc} with the mentioned substitutions being made, we call it a \emph{finitary disjunctive N-sequent calculus}.

For a finitary disjunctive N-sequent calculus $\bQ$ with disjunctive basis $(Q, R_{0})$ set 
\[
E = \CCC_{f}(Q, R_{0}), 
\]
define the token set $\Con$ as well as the entailment relation $\Vdash$ as in the preceding section, and let $\EEE_{\bQ} = (E, \Con, \Vdash, T)$.

\begin{pn}\label{pn-fseqinfw}
Let $\bQ$ be finitary disjunctive N-sequent calculus. Then $\EEE_{\bQ}$ is an algebraic L-information system with witnesses such that all tokens are reflexive and Condition~\ref{alg+} as well as the requirement that if $(a, X), (a, Y) \in \Con$ then $(a, X \cup Y) \in \Con$, hold.
\end{pn}

Note that in the present case the definition of the equivalence classes forming the elements of the Lindenbaum algebra $\LLL(\bQ)$ of $\bQ$ can be simplified. For $\phi \in \LLL_{f}(Q, R_{0})$ set
\[
[\phi]_{\bQ} = \set{\psi \in \LLL_{f}(Q, R_{0})}{\psi \dashv\vdash \phi}.
\] 

\begin{dn}\label{dn-fDlat}
Let $(L; 0, 1, \sqcap)$ be a meet-semilattice with least element 0 and greatest element 1.
\begin {enumerate}

\item\label{dn-fDlat-1} 
The semilattice is called \emph{finitary disjunctive semilattice} if every disjoint set $\{ x, y \}$ has a supremum $x \scup y$. 

\item\label{dn-fDlat-2} 
A finitary disjunctive semilattice $L$ is \emph{distributive} if 
\[
x \sqcap (y \scup z) = (x \sqcap y) \scup (x \sqcap z)
\]
 for all $x \in L$ and disjoint subsets $\{ y, z \}$ of $L$.
 
\item\label{dn-fDlat-3}
An element $u \in L \setminus \{ 0 \}$ of a distributive finitary disjunctive semilattice $L$ is a \emph{coprime} if, for any disjoint subset $\{ x, y \}$ of $L$, $u \sqsubseteq x \scup y$ implies that $u \sqsubseteq x$ or $u \sqsubseteq y$.

\item\label{dn-fDlat-4}
A distributive finitary disjunctive semilattice $L$ is \emph{finitely generated} if $1$ is a coprime and  for each $x \in L \setminus \{ 0 \}$, there is a nonempty finite disjoint set of coprimes of $L$ with supremum $x$.

\item\label{dn-fDlat-5}
A filter $\FFF$ of a finitely generated distributive finitary disjunctive semilattice $L$ is \emph{prime} if $x \scup y \in \FFF$ implies that $x \in \FFF$ or $y \in \FFF$, for every disjoint subset $\{ x, y \}$ of $L$. 

\end{enumerate}
\end{dn}

\begin{thm}[Wang, Li~\cite{wll20}]\label{thm-wl}
\begin{enumerate}

\item\label{thm-wl-1}
Let $\bQ$ be finitary disjunctive N-sequent calculus. Then the Lindenbaum algebra $\LLL(\bQ)$ of $\bQ$ is a finitely generated distributive finitary disjunctive semilattice where the lattice operations are lifted from their corresponding logical connectives, and thus the equivalence classes of irreducible simple conjunctions are exactly the coprimes of $\LLL(\bQ)$.

\item\label{thm-wl-2}
The set of a prime filters of a finitely generated distributive finitary disjunctive semilattice  forms a Lawson compact algebraic L-domain $\LL$ when ordered by set inclusion, and vice versa.

\end{enumerate}
\end{thm}

As analogue of Theorem~\ref{thm-maincj} we can now state the main result of this subsection.

\begin{thm}\label{thm-fmain}
Let $\bQ$ be a finitary disjunctive N-sequent calculus. Then an algebraic L-infor\-mation system $\EEE_{\bQ}$ with witnesses can be constructed such that:
\begin{enumerate}

\item\label{thm-fmain-1}
 All tokens are reflexive.
 
 \item\label{thm-fmain-2}
Condition~\ref{alg+} holds.

\item\label{thm-fmain-3}
If $(a, X), (a, Y) \in \Con$ then $(a, X \cup Y) \in \Con$ as well.

\item\label{thm-fmain-4}
The L-domain $|\EEE_{\bQ}|$ associated with $\EEE_{\bQ}$ is isomorphic to the L-domain  $\LL_{\bQ}$  generated by the Lindenbaum algebra $\LLL_{\bQ}$ of $\bQ$.
\end{enumerate}
\end{thm}

\subsection{Conjunctive sequent calculi}

Conjunctive sequent calculi have been introduced by Wang and Li~\cite{wlbc20} as a Gentzen-style logical calculus to represent bounded-complete continuous domains without greatest element. We will show in this section that each conjunctive sequent calculus defines an information system with witnesses  so that both generated isomorphic domains, and vice versa.

\begin{dn}\label{dn-cpf}
Let $P$ be a nonempty set $P$ and $T_{P} \in P$. Each element of $P$ is called  \emph{atomic formula}. The set $\LLL(P)$ of formulae is defined inductively as follows:
\begin{enumerate}
\item\label{dn-cpf-1}
Every atomic formula $p$ is in $\LLL(P)$.

\item\label{dn-cpf-2}
Whenever $\phi, \psi$ are in $\LLL(P)$, $\phi \wedge \psi$ is also in $\LLL(P)$.

\end{enumerate}  
\end{dn}

For any $\phi \in \LLL(P)$ we let $\overline{\phi}$ be the set of all atomic propositions $p$ occurring in $\phi$, and set $\overline{\Delta} = \set{p \in \overline{\phi}}{\phi \in \Delta}$, for any nonempty finite subset $\Delta$ of $\LLL(P)$. 

\begin{dn}\label{dn-csc}
A \emph{conjunctive sequent calculus} is a pair $(\LLL(P), \vdash_{P})$, where $\vdash_{P}$ is a relation on nonempty finite subsets of $\LLL(P)$ that is closed under the following derivation rules:

\[
\begin{array}{c@{\qquad\qquad}c}
\infer[(T_{p})]{\Gamma \vdash_{P} T_{P}}{} & \infer[\mathrm{(Weakening)}]{\Gamma', \Gamma \vdash_{P} \Delta}{\Gamma \vdash_{P} \Delta}  \\[3ex]
\infer=[(\wedge\mathrm{-left})]{\phi \wedge \psi, \Gamma \vdash_{P} \Delta}{\phi, \psi,\Gamma \vdash_{P} \Delta} & \infer=[\wedge\mathrm{-right}]{\Gamma \vdash_{P} \Delta, \phi \wedge \psi}{\Gamma \vdash_{P} \Delta, \phi & \Gamma \vdash_{P} \Delta, \psi} \\[3ex]
\multicolumn{2}{c}{\infer=[\mathrm{(Cut)}]{\Gamma \vdash_{P} \Delta}{\Gamma \vdash_{P} \phi & \phi \vdash_{P} \Delta}}.
\end{array}
\]
\end{dn}

For a conjunctive sequent calculus $(\LLL(P), \vdash_{P})$ a formula $\phi \in \LLL(P)$ is \emph{satisfiable}, if for some $\psi \in \LLL(P)$, $\phi \not\vdash_{P} \psi$. The set of all satisfiable formulae is denoted by $\SSS_{P}$. 

\begin{dn}\label{dn-cccons}
A conjunctive sequent calculus $(\LLL(P), \vdash_{P})$ is \emph{consistent}, if $P \subseteq \SSS_{P} \not= \LLL(P)$ and for every $\psi \in \SSS_{P}$ there is some $\phi \in \SSS_{P}$ with $\phi \vdash_{P} \psi$.
\end{dn}

Let $\SSb(P) = \set{\Gamma \fsubset \LLL(P)}{\Gamma \not= \emptyset \newand \bigwedge \Gamma \in \SSS_{P}}$ and define
 \[
 \Gamma \Vdash_{P} \Delta \Longleftrightarrow \Gamma \in \SSb_{P} \newand \Gamma \vdash_{P} \Delta.
 \]
 Moreover, for $X \subseteq \LLL(P)$ set
 \[
 X[\Vdash_{P}] = \set{\psi \in \LLL(P)}{\text{$\rho \Vdash_{P} \psi$ with $\overline{\rho} \subseteq \overline{\Gamma}$, for some nonempty $\Gamma \fsubset X$}}.
 \]
 Note that by \cite[Proposition~4.1]{wlbc20}, $X[\Vdash_{P}] = \bigwedge X[\Vdash_{p}]$, for nonempty finite subsets $X$ of $\LLL(P)$. 
 If $X = \{ \phi \}$, for some $\phi \in \SSS_{P}$, we write $\phi[\Vdash_{P}]$ instead of $\{ \phi \}[\Vdash_{P}]$. Note that in this case 
 \begin{equation}\label{eq-sfy}
 \phi[\Vdash_{P}] = \set{\psi \in \SSS_{P}}{\phi \Vdash_{P} \psi}.
 \end{equation}
 
 \begin{pn}\label{pn-ccinfw}
 Let $\PPP = (\LLL(P), \vdash_{P})$ be a consistent conjunctive sequent calculus. Then $\AAA_{\PPP} = (P, \Con, \vDash, T_{P})$ with
 $\Con = P \times (\set{\overline{\Gamma}}{\Gamma \in \SSb(P)} \cup \{ \emptyset \})$ and for $(p, X) \in \Con$ and $q \in P$,
 \[
 (p, X) \vDash q \Longleftrightarrow  \bigwedge X \Vdash_{P} q
 \]
 is an information system with witnesses the satisfies Condition~(\ref{bc}) such that the following statement hold:
 \begin{enumerate}
 
 \item\label{pn-ccinfw-1}
 $\Con$ is a proper subset of $P \times \PPP_{f}(P)$.
 
 \item\label{pn-ccinfw-1a}
For all $a \in A$ and $Y \fsubset  A$, if $(a, Y) \in \Con$, then there is some $X \fsubset A$ so that $(a, X) \in \Con$ and $(a, X) \vDash Y$.

 \item\label{pn-ccinfw-2}
  For $(p, X) \in \Con$ and $q \in P$,  $[X]_{p} = \bigwedge X[\Vdash_{P}] \cap P$.
 
 \item\label{pn-ccinfw-3}
 For  $\phi \in \SSS_{P}$ and $p \in P$, $\phi[\Vdash_{P}] = \set{\psi \in \SSS_{P}}{\overline{\psi} \subseteq [\overline{\phi}]_{p}}$.
 \end{enumerate}
 \end{pn}
 \begin{pf}
First we will verify the Conditions in Definition~\ref{dn-infsys} and (\ref{bc}). 

(\ref{dn-infsys-1}) holds as $P \subseteq \SSS_{P}$. (\ref{dn-infsys-2}) is a consequence of the consistency of $\PPP$. (\ref{dn-infsys-3}) follows with Rule ($T_{P}$), and (\ref{dn-infsys-4}) as in \cite[p.~6]{wlbc20}. (\ref{dn-infsys-5}) is obtained with Weakening and (\ref{dn-infsys-6}) with Cut.  (\ref{dn-infsys-7}-\ref{dn-infsys-9}) are obvious. For (\ref{dn-infsys-10}) assume for $(p, X) \in \Con$ that $(p, X) \vDash q$. Then $X \vdash_{P} q$. Apply the inverted cut rule and let $\phi$ be the cut formula. Set $Z = \overline{\phi}$ and let $r \in Z$. Then we have that $X \vdash_{P} \phi$ and thus $(p, X) \vDash (r, Z)$. Moreover, we have that $\phi \vdash_{P} Y$, which implies that $(r, Z) \vDash Y$. (\ref{bc}) is obvious again.

It remains to verify Statements~(\ref{pn-ccinfw-1}-\ref{pn-ccinfw-3}).
(\ref{pn-ccinfw-1}) Assume that $\Con = P \times \PPP_{f}(P)$ and let $\phi \in \LLL(P)$. Then $P \times \{ \overline{\phi} \} \in \SSb(P)$ and hence $\phi \in \SSS_{P}$, which contradicts the consistency of $\PPP$. 

(\ref{pn-ccinfw-1a}) is a consequence of consistency.

(\ref{pn-ccinfw-2}) Let $(p, X) \in \Con$ and $q \in P$. Then $\bigwedge X \in \SSS_{P}$ and 
\begin{align*}
q \in X[\Vdash_{P}] 
&\Longleftrightarrow \text{$\rho \Vdash_{P} q$, for some $\rho \in \SSS_{P}$ with $\overline{\rho} \subseteq X$} \\
&\Longleftrightarrow X \Vdash_{P} q \quad\text{(by Weakening and by letting $\rho = \bigwedge X$, respectively)} \\
&\Longleftrightarrow (p, X) \vDash q \\
&\Longleftrightarrow q \in [X]_{p}. 
\end{align*}

(\ref{pn-ccinfw-3}) Let $\phi \in \SSS_{P}$ and $p \in P$. Then it follows with~(\ref{eq-sfy}) that
\[
\psi \in \phi[\Vdash_{P}] 
\Longleftrightarrow \phi \Vdash_{P} \psi 
\Longleftrightarrow (p, \overline{\phi}) \vDash \overline{\psi}
\Longleftrightarrow \overline{\psi} \subseteq [\overline{\phi}]_{p}.
\]
\end{pf}

Note that Condition~(\ref{pn-ccinfw-1a}) above reverses Requirement~(\ref{dn-infsys-4}) in Definition~\ref{dn-infsys}.

As is shown in \cite[Theorem~4.1]{wlbc20}, $\set{\phi[\Vdash_{P}]}{\phi \in \SSS_{P}}$ is a basis of a bounded-complete continuous domain $\CC_{\PPP}$ without maximal element. $\set{[X]_{p}}{(p, X) \in \Con}$, on  the other hand, is a basis of the bounded-complete continuous domain $|\AAA_{P}|$. The relationship between both bases stated in the preceding proposition allows the construction of an isomorphism between both domains. So, we have the following result.

\begin{thm}\label{thm-csctoinf}
Let $\PPP = (\LLL(P), \vdash_{P})$ be a consistent conjunctive sequent calculus. Then an information system $\AAA_{\PPP} = (P, \Con, \vDash, T_{P})$ with witnesses can be constructed such that
\begin{enumerate}
\item\label{thm-csctoinf-1}
$\AAA_{\PPP}$ satisfies Condition~(\ref{bc}).

\item\label{thm-csctoinf-2}
$\Con$ is a proper subset of $P \times \PPP_{f}(P)$.

\item\label{thm-csctoinf-2a}
For all $a \in A$ and $Y \fsubset  A$, if $(a, Y) \in \Con$, then there is some $X \fsubset A$ so that $(a, X) \in \Con$ and $(a, X) \vDash Y$.

\item\label{thm-csctoinf-3}
The domain $|\AAA_{\PPP}|$ associated with $\AAA_{\PPP}$ is isomorphic to the bounded-complete continuous domain $\CC_{\PPP}$ generated by $\PPP$.

\end{enumerate}
\end{thm}

Let us next study the converse situation that an information system with witnesses satisfying Conditions~(\ref{thm-csctoinf-1}-\ref{thm-csctoinf-2a}) of the preceding result  is given. We will show that a consistent conjunctive sequent calculus can be derived from it so that both generate isomorphic domains.

\begin{pn}\label{pn-infcs}
Let $\AAA = (A, \Con, \vdash, \bt)$ be an information system with witnesses so that 
\begin{enumerate}
\item\label{pn-assinfcs-1}
Condition~(\ref{bc}) holds, 

\item\label{pn-assinfcs-2}
$\Con$ is a proper subset of $A \times \PPP_{f}(A)$, and

\item\label{pn-assinfcs-3}
For all $a \in A$ and $Y \fsubset  A$, if $(a, Y) \in \Con$, then there is some $X \fsubset A$ so that $(a, X) \in \Con$ and $(a, X) \vdash Y$.

\end{enumerate}
Moreover, let $\PPP_{\AAA} = (\LLL(P), \Vvdash_{P})$, where
$P= A$, $T_{p} = \bt$, and 
 for nonempty subsets $\Gamma, \Delta$ of $\LLL(P)$, 
\[
\Gamma \Vvdash_{P} \Delta,  
\]
if for some $a \in A$, $(a, \overline{\Gamma}) \in \Con$ and $(a, \overline{\Gamma}) \vdash \overline{\Delta}$, or for no $(a \in A)$, $(a, \overline{\Gamma}) \in  \Con$. Then $\PPP_{\AAA}$ is a consistent conjunctive sequent calculus such that the following statements hold:
\begin{enumerate}
\item\label{pn-infcs-1}
$\SSb(P) = \set{\Gamma \fsubset \LLL(P)}{\text{$\Gamma$ nonempty and for some $a \in A$, $(a, \overline{\Gamma}) \in \Con$}}$, 

\item\label{pn-infcs-2}
$[X]_{a} = \bigwedge X[\Vdash_{P}] \cap A$, for $(a, X) \in \Con$,

\item\label{pn-infcs-3}
$\phi[\Vdash_{P}] = \set{\psi \in \SSS_{P}}{\overline{\psi} \subseteq [\overline{\phi}]_{p}}$, for $\phi \in \SSS_{P}$ and $p \in P$.
\end{enumerate}
Here, $\Vdash_{P}$ is the restriction of $\Vvdash_{P}$ to $\SSb(P)$.
\end{pn}
\begin{pf}
We first show that $\Vvdash_{P}$ is closed under the rules in Definition~\ref{dn-csc}. Rule~($T_{P}$) is a consequence of Rules~\ref{dn-infsys}(\ref{dn-infsys-3},\ref{dn-infsys-5}). Weakening as well follows with \ref{dn-infsys}(\ref{dn-infsys-5}). Rule~($\wedge$-left) is obvious, as $\overline{\{ \phi, \psi \} \cup \Gamma} = \overline{\{ \phi \wedge \psi \} \cup \Gamma}$. Rule~($\wedge$-right) is obvious, if for no $a \in A$, $(a, \overline{\Gamma}) \in \Con$. In the other case it holds as $(a, \overline{\Gamma}) \vdash \overline{\Delta} \cup \overline{\phi}$ and $(a, \overline{\Gamma}) \vdash \overline{\Delta} \cup \overline{\psi}$ implies that $(a, \overline{\Gamma}) \vdash \overline{\Delta} \cup \overline{\phi \wedge \psi}$, and vice versa. The Cut rule follows with \ref{dn-infsys}(\ref{dn-infsys-6}), for its inversion use Rule~\ref{dn-infsys}(\ref{dn-infsys-10}): choose $\bigwedge Z$ as cut formula. The case that for no $a \in A$, $(a, \overline{\Gamma}) \in \Con$ is again obvious.

Let us next derive Statement~(\ref{pn-infcs-1}). Let $\phi \in \LLL(P)$ so that for some $a \in A$, $(a, \overline{\phi}) \in \Con$, and let $X \in \PPP_{f}(A)$ such that for no $a \in A$, $(a, X) \in \Con$. Set $\psi = \wedge X$. Assume that $\phi \Vvdash_{P} \psi$. Then $(a, \overline{\phi}) \vdash X$. Hence, $(a, X) \in \Con$, by Condition~\ref{dn-infsys}(\ref{dn-infsys-4}), a contradiction.
This shows that $\set{\phi \in \LLL(P)}{\text{$(a, \overline{\phi}) \in \Con$, for some $a \in A$}} \subseteq \SSS_{P}$. Because of Requirement~\ref{dn-infsys}(\ref{dn-infsys-1}), we moreover obtain that $P \subseteq \set{\phi \in \LLL(P)}{\text{$(a, \overline{\phi}) \in \Con$}}$.

Now, conversely, let $\phi \in \SSS_{P}$. Then there is some $\psi \in \LLL(P)$ with $\phi \not\Vvdash_{P} \psi$, which is only possible if the first case in the definition of $\Vvdash_{P}$ holds, that is that for some $a \in A$, $(a, \overline{\phi}) \in \Con$. Thus, $\SSS_{P} = \set{\phi \in \LLL(P)}{\text{$(a, \overline{\phi}) \in \Con$, for some $a \in A$}}$, which implies that Statement~(\ref{pn-infcs-1}) holds. With Assumption~(\ref{pn-assinfcs-2}) we thus obtain that $\PPP_{\AAA}$ is consistent.

The verification of Statements~(\ref{pn-infcs-2}) and (\ref{pn-infcs-3}) is easy and left to the reader.
\end{pf}

As in the preceding case it follows from Statements~(\ref{pn-infcs-2}) and (\ref{pn-infcs-3}) of the above propositions that the domains $|\AAA|$ and $\CC_{\PPP_{\AAA}}$, respectively, generated by the information system with witnesses $\AAA$ and its derived consistent conjunctive sequent calculus $\PPP_{\AAA}$ are isomorphic.

\begin{thm}\label{thm-inftocsc}
Let $\AAA = (A, \Con, \vdash, \bt)$ be an information system with witnesses such that 
\begin{enumerate}
\item\label{thm-assinftocsc-1}
Condition~(\ref{bc}) holds,

\item\label{thm-assinftocsc-2}
 $\Con$ is a proper subset of $A \times \PPP_{f}(A)$, and
 
 \item\label{thm-assinftocsc-3}
 For all $a \in A$ and $Y \fsubset  A$, if $(a, Y) \in \Con$, then there is some $X \fsubset A$ so that $(a, X) \in \Con$ and $(a, X) \vdash Y$.
 
 \end{enumerate}
Then a consistent conjunctive sequent calculus $\PPP_{\AAA}$ can be constructed so that the domain $\CC_{\PPP_{\AAA}}$ generated by $\PPP_{\AAA}$ is isomorphic to the domain $|\AAA|$ associated with  $\AAA$.
\end{thm}

\section{Conclusion}\label{sec-conc}

Scott-style information systems allow for a constructive approach to the theory of Scott domains, that is bounded-complete directed complete partial orders.  The present paper introduces two equivalent generalisations that both capture L-domains. 

The first of these is in the style of Kripke frames. Each node comes equipped with its own (local) logic. In addition there are rules for the consistent movement from one node to its successor. The given requirements are very natural: they are satisfied by any reasonable logic.

The other approach starts from the observation that in a Scott information system every consistent set of tokens possesses a canonical witness for its consistency.  In L-domains this property is no longer guaranteed to hold: canonical consistency witnesses do only exist relative to certain principal filters of the domain. Consequently, the consistency of a token set can only be stated by explicitly presenting its consistency witness.

Both approaches are very natural: Every continuous information system as introduced by Hoofman~\cite{ho93} and each (algebraic) information system as considered by Scott~\cite{sco82}, and/or Larsen/Winskel~\cite{lw84}, defines an information system with witnesses satisfying an additional characteristic condition. Conversely, for every information system with witnesses fulfilling the extra requirement a continuous (and/or algebraic) information system can be constructed so that both generate isomorphic bounded-complete domains.

From a proof-theoretical perspective, information systems generalise natural deduction style propositional logic. Other approaches to program logic such as Chen and Jung's disjunctive propositional logic~\cite{cj06} (see also Wang and Li~\cite{wll20} for the finitary subcase) as well as Wang and Li's conjunctive sequent calculi~\cite{wlbc20} are based on Gentzen's sequent calculus. Also for these calculi syntactic back-and-forth translations to information systems with witnesses have been presented such that the domains formed by the syntactic models (or theories) of the logics are preserved, up to isomorphism.

\section*{Acknowledgement}
The author is grateful to the referees for their useful remarks.

\end{document}